\documentclass[aj,twocolumn]{aastex631}

\usepackage{xspace}
\usepackage{comment}
\usepackage{subfigure}
\usepackage{soul}
\usepackage{multirow}

\newcommand{\TESS}{\emph{TESS}\xspace}
\newcommand{\toi}{TOI-4562\xspace}
\newcommand{\toib}{TOI-4562\,b\xspace}

\newcommand{\tcvalue}{1456.87168$^{+0.00101}_{-0.00101}$}
\newcommand{\tcfirstvalue}{1456.87991$^{+0.00142}_{-0.00120}$}
\newcommand{\tcsecondvalue}{1681.98324$^{+0.00136}_{-0.00142}$}
\newcommand{\tcthirdvalue}{2132.21577$^{+0.00088}_{-0.00095}$}
\newcommand{\tcfourthvalue}{2357.34544$^{+0.00079}_{-0.00080}$}
\newcommand{\tcfifthvalue}{2582.46321$^{+0.00142}_{-0.00036}$}
\newcommand{\pervalue}{225.11781$^{+0.00025}_{-0.00022}$}
\newcommand{\rprstarvalue}{0.09980$^{+0.00084}_{-0.00092}$}
\newcommand{\incvalue}{$89.06 \pm 0.04$}
\newcommand{\secoswvalue}{0.442$^{+0.110}_{-0.131}$}
\newcommand{\sesinwvalue}{0.752$^{+0.058}_{-0.063}$}
\newcommand{\mpvalue}{$0.0022 \pm 0.0005$}
\newcommand{\mStarvalue}{$1.192 \pm 0.057$} 

\newcommand{\loggvalue}{$4.39 \pm 0.01$}
\newcommand{\fehvalue}{$0.08 \pm 0.06$}
\newcommand{\agevalue}{$<$ 700}
\newcommand{\parallaxvalue}{2.85692$^{+0.01009}_{-0.00990}$}
\newcommand{\meanvalue}{5347$\pm13$}
\newcommand{\jitvalue}{$73^{+15}_{-15}$}
\newcommand{\ebvvalue}{$<0.01\, (3\sigma)$}
\newcommand{\rStarvalue}{$1.152 \pm 0.046$}  

\newcommand{\smarStarvalue}{147.4$^{+1.44}_{-1.26}$}
\newcommand{\smaAUvalue}{0.768$^{+0.005}_{-0.005}$}
\newcommand{\distvalue}{$350.0\pm1.2$}
\newcommand{\kampvalue}{106$^{+24}_{-26}$}
\newcommand{\mpjupvalue}{2.30$^{+0.48}_{-0.47}$}
\newcommand{\mpearthvalue}{732$^{+152}_{-149}$}
\newcommand{\rpjupvalue}{$1.118_{+0.013}^{-0.014}$}
\newcommand{\rpearthvalue}{$12.53 \pm 0.15$}
\newcommand{\densityvalue}{2.06$^{+0.42}_{-0.44}$}
\newcommand{\eccvalue}{0.76$^{+0.02}_{-0.02}$}
\newcommand{\wvalue}{60$^{+9}_{-8}$}
\newcommand{\bvalue}{0.60$^{+0.03}_{-0.04}$}
\newcommand{\tdepthvalue}{9961$^{+167}_{-184}$}
\newcommand{\tdurvalue}{$4.32 \pm 0.04$}
\newcommand{\teqvalue}{$318\pm4$}
\newcommand{\tperivalue}{717$^{+29}_{-37}$}
\newcommand{\tapovalue}{262$^{+3}_{-2}$}
\newcommand{\sepvalue}{35}
\newcommand{\teffCHIRON}{$6096 \pm 50$}
\newcommand{\loggCHIRON}{$4.4 \pm 0.1$}
\newcommand{\fehCHIRON}{$0.1 \pm 0.1$}
\newcommand{\vsiniCHIRON}{$17 \pm 0.5$}
\newcommand{\vsiniFEROS}{$15.7 \pm 0.5$}
\newcommand{\Protvalue}{$3.86_{-0.08}^{+0.05}$}
\newcommand{\ProtvalueASAS}{$3.84 \pm 0.02$}

\newcommand{\teff}{\ensuremath{T_{\rm eff}}}
\newcommand{\met}{[$\mathrm{M}$/H]}
\newcommand{\vsini}{$v$~sin $i$}
\newcommand{\kms}{\,km\,s$^{-1}$}
\newcommand{\ms}{\,m\,s$^{-1}$}

\newcommand{\degree}{$^{\circ}$}
\newcommand{\masyr}{mas\,yr$^{-1}$}
\newcommand{\Bmag}{$B$}
\newcommand{\Vmag}{$V$}
\newcommand{\Tmag}{$T$}
\newcommand{\gaiaG}{$G$}
\newcommand{\gaiaBP}{$B_{\mathrm{p}}$}
\newcommand{\gaiaRp}{$R_{\mathrm{p}}$}
\newcommand{\Jmag}{$J$}
\newcommand{\Hmag}{$H$}
\newcommand{\Kmag}{$K_{\mathrm{s}}$}
\newcommand{\Wi}{$W_{\mathrm{1}}$}
\newcommand{\Wii}{$W_{\mathrm{2}}$}
\newcommand{\Wiii}{$W_{\mathrm{3}}$}
\newcommand{\Rstar}{$R_{\mathrm{\star}}$}
\newcommand{\Mstar}{$M_{\mathrm{\star}}$}

\newcommand{\rhostar}{$\rho_{\mathrm{\star}}$}
\newcommand{\rhocirc}{$\rho_{\mathrm{circ}}$}

\newcommand{\Msun}{$M_{\odot}$}
\newcommand{\Rsun}{$R_{\odot}$}
\newcommand{\TBJD}{$BJD_{TDB}$}
\newcommand{\REarth}{$R_{\oplus}$}
\newcommand{\MEarth}{$M_{\oplus}$}
\newcommand{\RJup}{$R_{\mathrm{J}}$}
\newcommand{\MJup}{$M_{\mathrm{J}}$}
\newcommand{\Porb}{$P_{\mathrm{orb}}$}
\newcommand{\Prot}{$P_{\star}$}
\newcommand{\Mp}{$M_{\mathrm{p}}$}
\newcommand{\Rp}{$R_{\mathrm{p}}$}
\newcommand{\RprStar}{$R_{\mathrm{p}}$/$R_{\mathrm{\star}}$}
\newcommand{\ars}{$a$/$R_{\mathrm{\star}}$}
\newcommand{\tc}{$T_{\mathrm{c}}$}
\newcommand{\secosw}{$\sqrt{e}\cos\omega$}
\newcommand{\sesinw}{$\sqrt{e}\sin\omega$}
\newcommand{\offsetchiron}{$\gamma_{\mathrm{rel}}$}

\newcommand{\Kamp}{$K_{\mathrm{amp}}$}

\shorttitle{The young \toib}
\shortauthors{Heitzmann et al.}

\newcommand{\USQ}{University of Southern Queensland, Centre for Astrophysics, West Street, Toowoomba, QLD 4350 Australia}
\newcommand{\PennStateAA}{Department of Astronomy \& Astrophysics, 525 Davey Laboratory, The Pennsylvania State University, University Park, PA, 16802, USA}
\newcommand{\PennStateCEHW}{Center for Exoplanets and Habitable Worlds, 525 Davey Laboratory, The Pennsylvania State University, University Park, PA, 16802, USA}
\newcommand{\CCA}{Center for Computational Astrophysics, Flatiron Institute, 162 Fifth Avenue, New York, NY 10010, USA}
\newcommand{\CfA}{Center for Astrophysics \textbar \ Harvard \& Smithsonian, 60 Garden Street, Cambridge, MA 02138, USA}
\newcommand{\Caltech}{Cahill Center for Astrophysics, California Institute of Technology, Pasadena, CA 91125, USA}
\newcommand{\IRAP}{Institut de Recherche en Astrophysique et Plan\'{e}tologie, Universit\'{e} de Toulouse, CNRS, CNES, 14 avenue Edouard Belin, 31400 Toulouse, France}
\newcommand{\LiegeUni}{Astrobiology Research Unit, Université de Liège, 19C Allée du 6 Août, 4000 Liège, Belgium}
\newcommand{\MITAtmo}{Department of Earth, Atmospheric and Planetary Science, Massachusetts Institute of Technology, 77 Massachusetts Avenue, Cambridge, MA 02139, USA}
\newcommand{\IACSpain}{Instituto de Astrof\'isica de Canarias (IAC), Calle V\'ia L\'actea s/n, 38200, La Laguna, Tenerife, Spain}
\newcommand{\QMU}{Astronomy Unit, Queen Mary University of London, Mile End Road, London E1 4NS, UK}
\newcommand{\CAM}{Astrophysics Group, Cavendish Laboratory, J.J. Thomson Avenue, Cambridge CB3 0HE, UK}

\newcommand{\AIUniChile}{Facultad de Ingeniería y Ciencias, Universidad Adolfo Ib\'a\~nez, Av.\ Diagonal las Torres 2640, Pe\~nalol\'en, Santiago, Chile}
\newcommand{\MIAChile}{Millennium Institute of Astrophysics MAS, Nuncio Monsenor Sotero Sanz 100, Of. 104, Providencia, Santiago, Chile}
\newcommand{\MaxPlank}{Max Planck Institute for Astronomy, K\"onigstuhl 17, D-69117—Heidelberg, Germany}
\newcommand{\KeeleUni}{Astrophysics Group, Keele University, Staffordshire, ST5 5BG, UK}
\newcommand{\kavlimit}{Department of Physics and Kavli Institute for Astrophysics and Space Research, Massachusetts Institute of Technology, Cambridge, MA 02139, USA}
\newcommand{\NESI}{NASA Exoplanet Science Institute, Caltech/IPAC, 1200 E. California Ave, Pasadena, CA 91125, USA}
\newcommand{\WisconsinMadisonUni}{Department of Astronomy, University of Wisconsin-Madison, Madison, WI 53706, USA}
\newcommand{\UTAustin}{Department of Astronomy, The University of Texas at Austin, Austin, TX 78712, USA}
\newcommand{\plahunters}{Citizen Scientist, Zooniverse c/o University of Oxford, Keble Road, Oxford OX1 3RH, UK}

\begin{document}

\title{\toib: A highly eccentric temperate Jupiter analog orbiting a young field star.}
\correspondingauthor{Alexis Heitzmann}
\email{alexis.heitzmann@usq.edu.au}

\author[0000-0002-8091-7526]{Alexis Heitzmann}
\affiliation{\USQ}

\author[0000-0002-4891-3517]{George Zhou} 
\affiliation{\USQ}

\author[0000-0002-8964-8377]{Samuel N.~Quinn} 
\affiliation{\CfA}

\author[0000-0003-0918-7484]{Chelsea X. Huang} 
\affil{\USQ}

\author[0000-0002-3610-6953]{Jiayin Dong} 
\altaffiliation{Flatiron Research Fellow}
\affiliation{\CCA}
\affiliation{\PennStateAA}
\affiliation{\PennStateCEHW}

\author[0000-0002-0514-5538]{L. G. Bouma} 
\altaffiliation{51 Pegasi b Fellow}
\affiliation{\Caltech}

\author[0000-0001-9677-1296]{Rebekah I. Dawson} 
\affiliation{\PennStateAA}
\affiliation{\PennStateCEHW}

\author[0000-0001-5522-8887]{Stephen C.~Marsden} 
\affiliation{\USQ}

\author[0000-0001-7294-5386]{Duncan Wright} 
\affiliation{\USQ}

\author[0000-0001-7624-9222]{Pascal Petit} 
\affiliation{\IRAP}

\author[0000-0001-6588-9574]{Karen A.\ Collins} 
\affiliation{\CfA}

\author[0000-0003-1464-9276]{Khalid Barkaoui} 
\affiliation{\LiegeUni}
\affiliation{\MITAtmo}
\affiliation{\IACSpain}

\author[0000-0001-9957-9304]{Robert A.~Wittenmyer} 
\affil{\USQ}


\author[0000-0003-2851-3070]{Edward Gillen} 
\altaffiliation{Winton Fellow}
\affiliation{\QMU}
\affiliation{\CAM}


\author[0000-0002-9158-7315]{Rafael Brahm} 
\affiliation{\AIUniChile}
\affiliation{\MIAChile}

\author[0000-0002-5945-7975]{Melissa Hobson} 
\affiliation{\MaxPlank}
\affiliation{\MIAChile}

\author[0000-0002-3439-1439]{Coel Hellier} 
\affiliation{\KeeleUni}

\author[0000-0002-0619-7639 ]{Carl Ziegler}
\affiliation{Department of Physics, Engineering and Astronomy, Stephen F. Austin State University, 1936 North St, Nacogdoches, TX 75962, USA}

\author[0000-0001-7124-4094]{C\'{e}sar Brice\~{n}o}
\affiliation{Cerro Tololo Inter-American Observatory, Casilla 603, La Serena, Chile}

\author{Nicholas Law}
\affiliation{Department of Physics and Astronomy, The University of North Carolina at Chapel Hill, Chapel Hill, NC 27599-3255, USA}

\author[0000-0003-3654-1602]{Andrew W. Mann}
\affiliation{Department of Physics and Astronomy, The University of North Carolina at Chapel Hill, Chapel Hill, NC 27599-3255, USA}

\author[0000-0002-2532-2853]{Steve B. Howell}
\affil{NASA Ames Research Center, Moffett Field, CA 94035, USA}

\author[0000-0003-2519-6161]{Crystal~L.~Gnilka}
\affil{NASA Ames Research Center, Moffett Field, CA 94035, USA}

\author[0000-0001-7746-5795]{Colin Littlefield}
\affiliation{Bay Area Environmental Research Institute, Moffett Field, CA 94035, USA}
\affiliation{NASA Ames Research Center, Moffett Field, CA 94035, USA}

\author[0000-0001-9911-7388]{David W.~Latham} 
\affiliation{\CfA}

\author[0000-0001-6513-1659]{Jack J. Lissauer} 
\affiliation{Space Science \& Astrobiology Division
, MS 245-3, NASA Ames Research Center, Moffett Field, CA 94035, USA}

\author[0000-0003-4150-841X]{Elisabeth R.~Newton} 
\affiliation{Department of Physics and Astronomy, Dartmouth College, Hanover, NH 03755, USA}

\author[0000-0001-9626-0613]{Daniel M.~Krolikowski} 
\affiliation{\UTAustin}

\author[0000-0002-6549-9792]{Ronan Kerr} 
\affiliation{\UTAustin}

\author[0000-0001-7337-5936]{Rayna Rampalli} 
\affiliation{Department of Physics and Astronomy, Dartmouth College, Hanover, NH 03755, USA}

\author[0000-0001-7371-2832]{Stephanie T. Douglas}  
\affiliation{Department of Physics, Lafayette College, 730 High St., Easton, PA 18042, USA}

\author[0000-0002-9138-9028]{Nora L.~Eisner} 
\affiliation{Sub-department of Astrophysics, University of Oxford, Keble Rd, Oxford, United Kingdom}

\author{Nathalie Guedj}
\affiliation{\plahunters}

\author{Guoyou Sun} 
\affiliation{\plahunters}

\author{Martin Smit} 
\affiliation{\plahunters}

\author{Marc Huten} 
\affiliation{\plahunters}

\author{Thorsten Eschweiler} 
\affiliation{\plahunters}

\author{Lyu Abe}
\affiliation{Universit\'e C\^ote d’Azur, Observatoire de la C\^ote d’Azur, CNRS,
Laboratoire Lagrange, CS 34229, F-06304 Nice Cedex 4, France}

\author[0000-0002-7188-8428]{Tristan Guillot}
\affiliation{Universit\'e C\^ote d’Azur, Observatoire de la C\^ote d’Azur, CNRS,
Laboratoire Lagrange, CS 34229, F-06304 Nice Cedex 4, France}


\author[0000-0003-2058-6662]{George Ricker}  
\affiliation{Department of Physics and Kavli Institute for Astrophysics and Space Research, Massachusetts Institute of Technology, Cambridge, MA 02139, USA}

\author[0000-0001-6763-6562]{Roland Vanderspek} 
\affiliation{Department of Physics and Kavli Institute for Astrophysics and Space Research, Massachusetts Institute of Technology, Cambridge, MA 02139, USA}

\author[0000-0002-6892-6948]{Sara Seager}
\affil{Department of Earth, Atmospheric, and Planetary Sciences, Massachusetts Institute of Technology, Cambridge, MA 02139, USA}
\affil{Department of Physics and Kavli Institute for Astrophysics and Space Research, Massachusetts Institute of Technology, Cambridge, MA 02139, USA}
\affil{Department of Aeronautics and Astronautics, Massachusetts Institute of Technology, Cambridge, MA 02139, USA}

\author[0000-0002-4715-9460]{Jon M. Jenkins} 
\affiliation{NASA Ames Research Center, Moffett Field, CA 94035, USA}

\author[0000-0002-8219-9505]{Eric B. Ting} 
\affiliation{NASA Ames Research Center, Moffett Field, CA 94035, USA}

\author[0000-0002-4265-047X]{Joshua N.~Winn} 
\affiliation{Department of Astrophysical Sciences, Princeton University, Princeton, NJ 08544, USA}


\author[0000-0002-5741-3047]{David R.~Ciardi} 
\affiliation{\NESI}

\author[0000-0001-7246-5438]{Andrew M.~Vanderburg} 
\affiliation{\kavlimit}
\affiliation{\WisconsinMadisonUni}

\author[0000-0002-7754-9486]{Christopher~J.~Burke}
\affiliation{Department of Physics and Kavli Institute for Astrophysics and Space Research, Massachusetts Institute of Technology, Cambridge, MA 02139, USA}

\author[0000-0003-1286-5231]{David~R.~Rodriguez}
\affiliation{Space Telescope Science Institute, 3700 San Martin Drive, Baltimore, MD, 21218, USA}

\author[0000-0002-6939-9211]{Tansu~Daylan}
\affiliation{Department of Astrophysical Sciences, Princeton University, Princeton, NJ 08544, USA}
\affiliation{LSSTC Catalyst Fellow}

\begin{abstract}
We report the discovery of \toib (TIC-349576261), a Jovian planet orbiting a young F7V-type star, younger than the Praesepe/Hyades clusters (\agevalue~Myr)
. This planet stands out because of its unusually long orbital period for transiting planets with known masses (\Porb~= \pervalue~days), and because it has a substantial eccentricity ($e$~=~\eccvalue). The location of \toi near the southern continuous viewing zone of \TESS allowed observations throughout 25 sectors, enabling an unambiguous period measurement from \TESS alone. Alongside the four available \TESS transits, we performed follow-up photometry using the South African Astronomical Observatory node of the Las Cumbres Observatory, and spectroscopy with the CHIRON spectrograph on the 1.5 m SMARTS telescope. We measure a radius of \rpjupvalue~\RJup, and a mass of \mpjupvalue~\MJup\,for \toib. The radius of the planet is consistent with contraction models describing the early evolution of the size of giant planets. We detect tentative transit timing variations at the $\sim$\,20 min level from five transit events, favouring the presence of a companion that could explain the dynamical history of this system if confirmed by future follow-up observations. With its current orbital configuration, tidal timescales are too long for \toib to become a hot-Jupiter via high eccentricity migration, though it is not excluded that interactions with the possible companion could modify \toib's eccentricity and trigger circularization. The characterisation of more such young systems is essential to set constraints on models describing giant planet evolution.
\end{abstract}


\section{Introduction} \label{sec:intro}

Planetary systems evolve rapidly within the first hundreds of millions of years of formation. The architectures of the systems evolve before settling into their eventual orbital configuration. Planets with extensive gaseous envelopes are expected to undergo contraction and cooling and experience observable changes in radius within this time frame. Observations of planets around young stars help anchor our understanding of this era of rapid change and help define models of planet formation and evolution. In particular, Jovian planets in distant orbits are less affected by stellar irradiation than close-in hot Jupiters. Transiting cold Jupiters around young stars can therefore provide constraints for cooling and contraction of giant planet evolution models. The orbital properties of these planets can also help to narrow down the timescales of dynamical evolution experienced by many other giant planets discovered to date. 

Numerous mechanisms are responsible for the formation and evolution of close-in Jovian planets. These mechanisms vary by the distribution of planets that they produce and by the timescales at which they operate. We can best assess the prevalence of these multiple formation channels via a census of the gas giant population as a function of time (see \citealt{2018ARA&A..56..175D}). Such a temporal survey of planetary systems can unveil the roles that in-situ formation (review in~\citealt{2014prpl.conf..619C}), disk migration (review in~\citealt{2014prpl.conf..667B}) and high eccentricity migration (review in~\citealt{2018ARA&A..56..175D}) played in shaping our current gas giant population. For example, planets can gravitationally interact with their depleting gas disks, resulting in moderately eccentric final orbits within a few million years \citep[e.g.,][]{2003ApJ...586.1374N, 2015ApJ...812...94D,2021MNRAS.500.1621D}. On the other extreme, excitation via stellar fly-bys can occur on the hundreds of millions of years timescale \citep[e.g.,][]{2016ApJ...816...59S}.

Gas giants also undergo significant contraction in the first hundred million years post formation. In models, the rate of contraction is strongly dependent on the initial conditions of the planet post formation, such as their envelope-core mass ratio and initial luminosities \citep[e.g.,][]{2007ApJ...659.1661F,2019A&A...623A..85L}. It is clear, however, that the radius distribution of close-in Jovian planets is shaped by external factors that retard their contraction \citep[e.g.,][]{2002A&A...385..156G,2003A&A...402..701B,2010ApJ...714L.238B}. Young planets in distant orbits provide simpler key tests for gas giant evolution. 

Missions like Kepler, K2, and the Transiting Exoplanet Survey Satellite (\TESS; \citealt{2015JATIS...1a4003R}) have brought forth a growing number of planetary systems about young stars \citep[e.g.,][]{2019ApJ...880L..17N,2020AJ....160..179M,2020Natur.582..497P, 2022AJ....163..121B,2022arXiv220501112B, 2022AJ....163..289Z}. However, true young Jovian analogues are rare. Interestingly, \citet{2021NatAs...6..232S} measured the masses of the giant planets in the 22 Myr old V1298 Tau system \citep{2019ApJ...885L..12D,2019AJ....158...79D}, finding that the two Jovian planets have already settled to their expected final radii, a process that is predicted to take hundreds of millions of years by contraction models. Other close-in Jovian-sized planets have also been found around young stars \citep{2020AJ....160...33R,2020AJ....160..239B,2021arXiv211009531M}, but strong stellar activity has yet prevented their mass from being measured. 

We report the discovery of a young transiting Jovian planet in a distant orbit around a \agevalue~Myr old star. \toi hosts a temperate-Jupiter in a 225\,day period orbit near the \TESS continuous viewing zone. Along with additional observations from our ground-based photometric follow-up campaign, five total transits of the planet were obtained, unambiguously identifying the period of the system. Radial velocity monitoring over the following two years provided a mass and eccentricity measurement for the young planet. In addition, data from FEROS helped to constrain the stellar parameters, and high resolution images from Gemini-South and SOAR helped to rule out false positive scenarios, confirming the transit candidate as a true planet. We also constrained the age of \toi via gyrochronology and lithium. Finally we detect a transit timing variations (TTV) signature, indicative of a perturbing companion in the system. \toib is one of the longest period transiting temperate Jupiters discovered by \TESS, and the youngest amongst such planets. Missions like \TESS and \emph{PLATO}~\citep{2014ExA....38..249R} have the potential to uncover this special population that critically constrains cooling models and migration pathways for Jovian planets.

\begin{figure*}
\centering
 \includegraphics[width=0.8\textwidth]{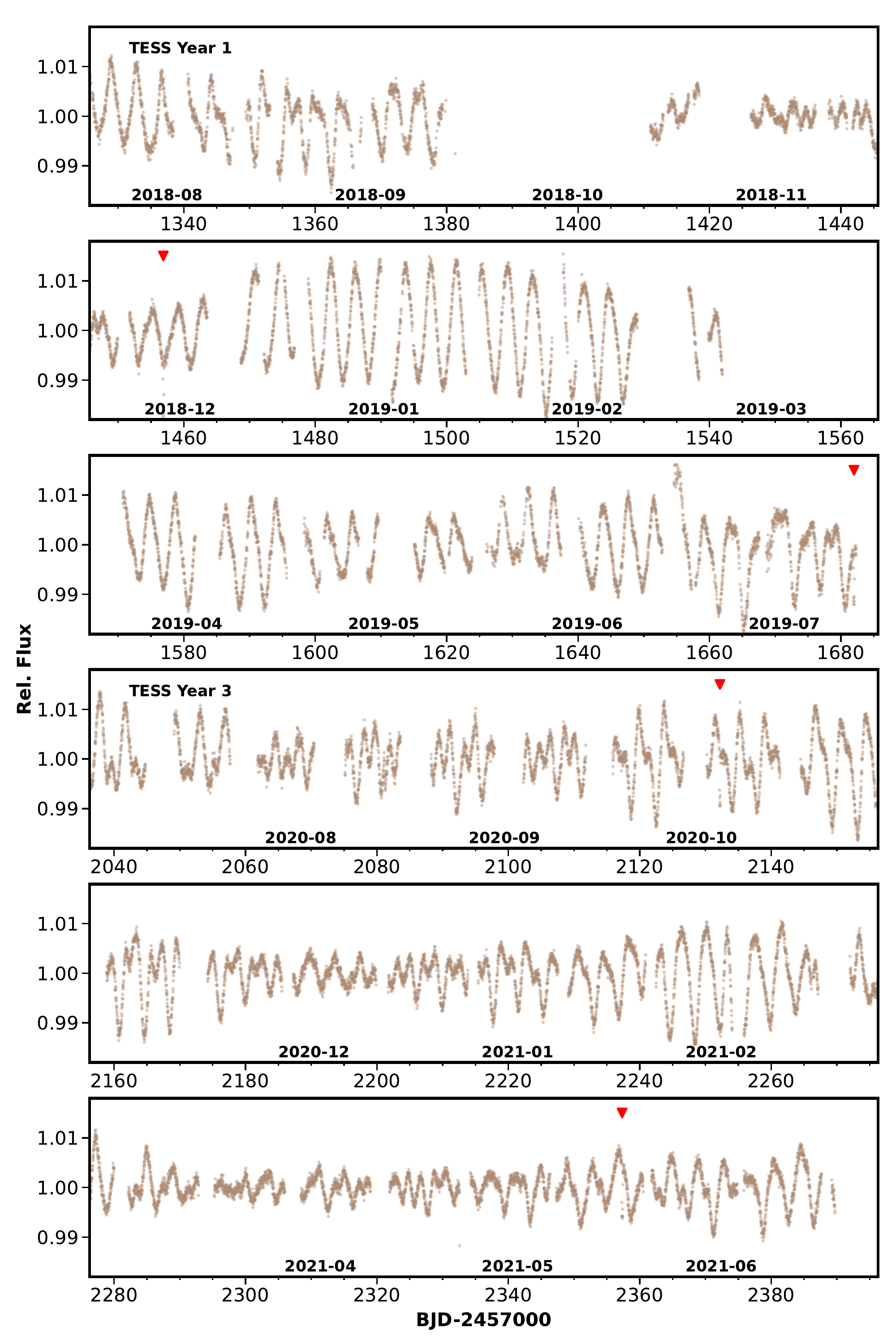}
\caption{\TESS lightcurve (brown) of \toib from all 25 available sectors. Photometry prior to sector 13 were obtained at 30 minute cadence, while latter observations were obtained at 2 minute cadence, and binned in this figure to 30 min for clarity. The four \toib transits from Sectors 5, 13, 30, and 38 are marked by red arrows. The host star exhibits up to $\sim$3\% peak-to-peak stellar rotational modulation due to its youth. }\label{fig:full_lc}
\end{figure*}

\section{Observations} \label{sec:obs}

\subsection{\TESS: Photometry} \label{subsec:tess}

The transiting planet candidate around \toi was first identified from observations by \TESS. \toi lies in the Southern Continuous Viewing Zone of \TESS, and therefore received near-uninterrupted photometric monitoring during years 1 and 3 of operations. The target received observations at 30\,minute cadence during Sectors 1-8 (2018-07-25 to 2019-02-28) and 10-13 (2019-03-26 to 2019-07-18), and 2\,minute target-pixel-stamp observations during sectors 27-39 (2020-07-04 to 2021-06-24).
The transit signature of TOI-4562b was detected by the TESS Science Processing Operations Center (SPOC;~\cite{2016SPIE.9913E..3EJ}) at NASA Ames Research Center during a transit search of sectors 27 through 39 with an adaptive, noise-compensating matched filter~\citep{2002ApJ...575..493J,2010SPIE.7740E..0DJ,2020TPSkdph}. The transit signature passed all the diagnostic tests in the Data Validation report~\citep{Twicken:DVdiagnostics2018} and was fitted with an initial limb-darkened transit model~\citep{Li:DVmodelFit2019}. In particular, the transit signal passed the difference image centroiding test, which localized the source of the transits to within $1.0 \pm 2.5$ arcsec. The TESS Science Office reviewed the diagnostic information and released an alert to the community for TOI-4562b on 28 October 2021 \citep{guerrero:TOIs2021ApJS}.

We make use of the MIT Quicklook pipeline \citep{2020RNAAS...4..204H} photometric extraction from the Full Frame Image observations. In addition, where available, we make use of the 2\,minute cadence target pixel file observations from the crowding and flux fraction corrected Simple Aperture Photometry (CROWDSAP) light curves \citep{twicken:PA2010SPIE,morris2020} made available by SPOC. Because of the large stellar variability seen in the light curve, we used the SAP light curves rather than the Pre-search Data Conditioning SAP (PDCSAP) flux and performed the detrending using a high order spline interpolation \citep{2014PASP..126..948V}

\begin{figure*}
\centering
 \includegraphics[width=0.8\textwidth]{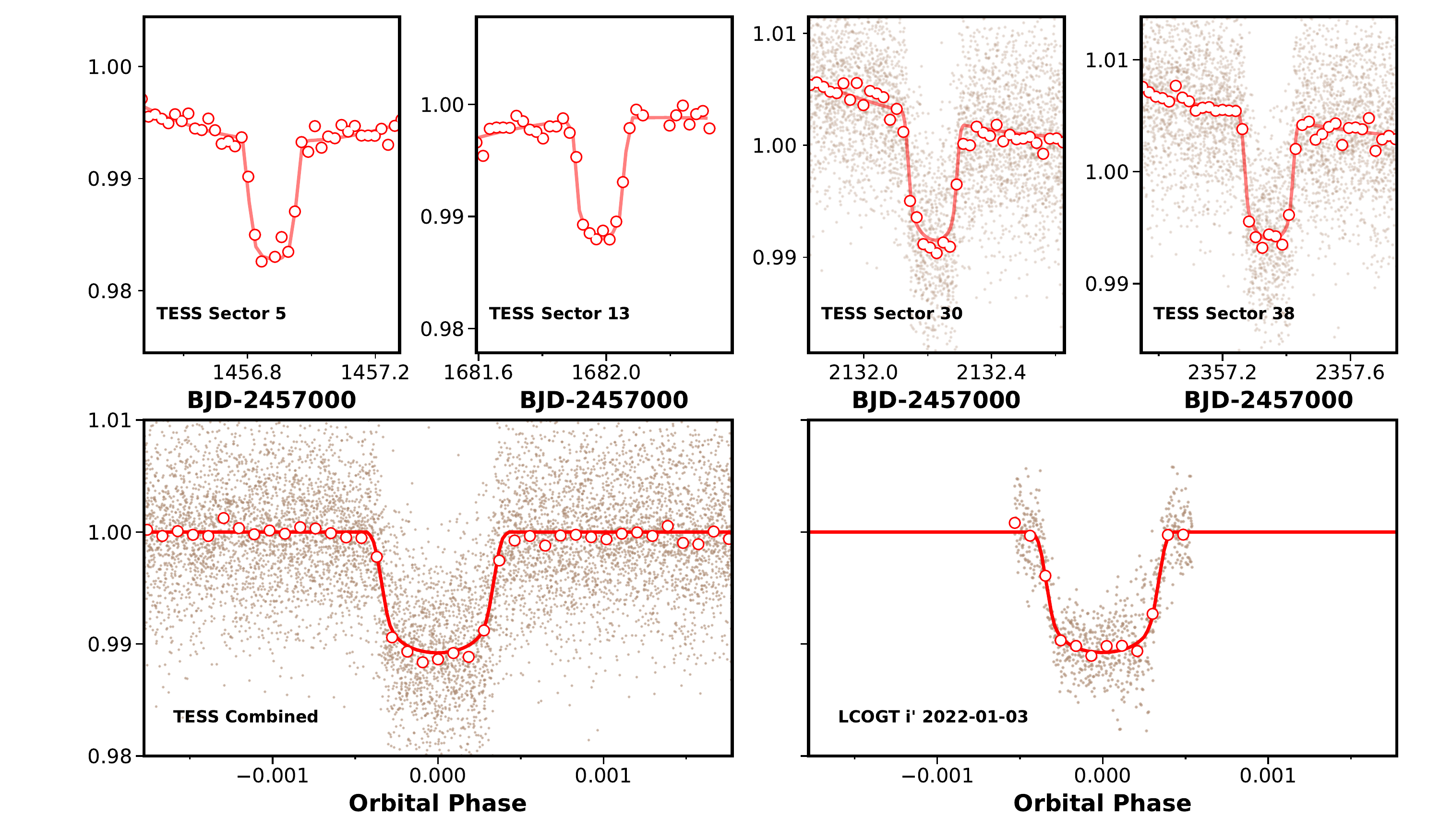}
\caption{\textbf{Top} Individual transits from \TESS from Sectors 5 and 13 at 30 minute cadence, and from Sectors 30 and 38 at 2\,minute cadence. The red circles indicate the measured data for sectors 5 and 13, and the binned data at 30 minute cadence for Sectors 30 and 38. The 2-minute cadence data for Sectors 30 and 38 are plotted as brown points. The best fit model, incorporating the transit timing variations in Section~\ref{sec:analysis}, and out-of-transit trends, are shown by the red lines. \textbf{Bottom left} The phase folded \TESS transit and best fit model. \textbf{Bottom right} The combined follow-up LCOGT 1\,m observations from 2022-01-03 in $i'$ band.}
\label{fig:transits}
\end{figure*}

The full \TESS light curve covering all sectors of observations is presented in Figure~\ref{fig:full_lc}. During the two (non-consecutive) years of near-continuous observations a total of 4 transits were captured by \TESS. Figure~\ref{fig:transits} shows the zoomed in region around each of these transits.  

\toi was first identified as a potential young star due to its strong rotational modulation \citep{2021AJ....161....2Z}, as part of our program to survey for planets around young field stars. We performed a search for transiting signals around \toi via a Box-least-squares period search \citep{2002A&A...391..369K} after removal of the stellar modulation signal with the splines. This detrending was not the one used for the transit modeling, described in section~\ref{subsec:transits}. 


\subsection{Follow-up photometry} \label{subsec:LCO}

We obtained follow-up photometric confirmation of the planetary transit via the Las Cumbres Observatory Global Network \citep[LCOGT;][]{2013PASP..125.1031B}. Transit opportunities for a 225\,day period planet are rare from the ground (see Table~\ref{tab:upcoming_transits}). We captured the full transit of \toib on 2022-01-03 UTC from the South African Astronomical Observatory (SAAO) node of LCOGT via two 1\,m telescopes. The observations were obtained with the \emph{Sinistro} 4K$\times$4K cameras in the Sloan $i'$ filter. The observations were calibrated via the \textsc{BANZAI} pipeline \citep{McCully:2018}, and light curves were extracted via the \textsc{AstroImageJ} package \citep[AIJ;][]{2017AJ....153...77C} using circular apertures with radius $4\farcs7$, which exclude flux from all known nearby Gaia EDR3 and TESS Input Catalog stars.  The combined light curves (after removing systematics) and best fit model are shown in Figure~\ref{fig:transits}.

In addition, a transit on 2022-08-16 was attempted from the SAAO node of LCOGT via one 1\,m telescope, as well as the Antarctica Search for Transiting ExoPlanets (ASTEP) facility \citep{2015AN....336..638G,2016MNRAS.463...45M}, located at the East Antarctic plateau. A 25\,minute segment was captured out of transit, but no portions of a transit event was recorded, and the dataset not included in the modeling presented below. 

\subsection{CHIRON/SMARTS: Spectroscopy} \label{subsec:chiron}

To characterize the radial velocity orbit of \toib and constrain the properties of the host star, we obtained 84 spectroscopic observations of \toi using the CHIRON facility. To capture the long orbital period of \toib, the velocities spanned two observing seasons, from 2020-12-09 to 2022-01-23; the resulting radial velocities are given in Table~\ref{tab:RVs}. CHIRON is a fiber-fed high resolution echelle spectrograph on the 1.5\,m SMARTS telescope at Cerro Tololo Inter-American Observatory, Chile \citep{2013PASP..125.1336T}. Due to the faintness of the host star, spectral observations were obtained in the `fiber' mode of CHIRON, yielding a resolving power of $R\sim28,000$ over the wavelength range of 4100 to $8700$\,\AA{}, and an average signal-to-noise of $\sim 100$ per resolution element at the Mg b line wavelength region. 

We make use of the extracted spectra from the standard CHIRON pipeline described in \citet{2021AJ....162..176P}. Radial velocities were derived from the observations via a least-squares deconvolution against a non-rotating ATLAS9 spectral template \citep{Castelli:2004}. The resulting broadening profile is fitted via a kernel describing the effects of radial velocity shift, rotational, macroturbulent, and instrumental broadening. The derived velocities are presented in Table~\ref{tab:RVs} and shown in Figure~\ref{fig:RV}.

To estimate the spectroscopic properties of the host star, we matched each spectrum against an observed library of $\sim 10,000$ spectra pre-classified by the Spectroscopic Classification Pipeline \citep{2012Natur.486..375B}. The matching was performed by first training the pre-classified library via a gradient boosting classifier using \textsc{scikit-learn}, and then classifying the observed spectrum. We found that \toi has an effective temperature of \teff\,=\,\teffCHIRON~K, a surface gravity of log\,$g$\,=\,\loggCHIRON~dex, a bulk metallicity of \met\,=\,\fehCHIRON~dex and a line of sight projected stellar rotational velocity of \vsini\,=\,\vsiniCHIRON~\kms. Since the CHIRON dataset overwhelms the other datasets we obtained for TOI-4562 in quantity, we adopt these parameters as Gaussian priors in the global analysis of the system described in Section~\ref{sec:analysis}. We note a general consensus between the spectral parameters from CHIRON and those presented below in Section~\ref{subsec:FEROSGALAH}.

We also check for the possibility that the velocity variations we observe are due to a spectroscopically blended companion rather than the host star. We compare the broadening measured from the line profiles against the velocities and find no correlation. If a blended companion is causing the radial velocity offset, then the line profiles should be broadest at the orbital quadratures, and narrowest at conjunctions. We therefore find no evidence that the velocity variations originate from a blended companion.

\subsection{FEROS \& GALAH: Spectroscopy} \label{subsec:FEROSGALAH}

The FEROS spectrograph, attached to the MPG 2.2 m \citep{1999Msngr..95....8K} telescope at La Silla Observatory, gathered 11 spectra of \toi. Spectra are co-added, with a signal to noise ratio per spectra ranging between 52 and 82, and atmospheric parameters are derived using \textsc{ZASPE} \citep{2017MNRAS.467..971B}. We find \teff\,=\,6280\,$\pm$\,100 K, log $g$\,= $4.49 \pm 0.10$, \met\,=\,$0.24 \pm 0.05$~dex and \vsini\,=\,\vsiniFEROS~\kms. We chose not to include the FEROS data in the RV modelling. All points fall near phases (-0.4, 0.025 and 0.35) where the RV signal is close to 0 and therefore don't meaningfully contribute, while adding one instrument and the associated extra
parameters. Using the \textsc{ceres} pipeline \citep{2017PASP..129c4002B}, we also recover chromospheric emission indices, tracers of stellar activity. The core emission of the $\mathrm{H_{\alpha}}$ line at 6562.808 \r{A} is $\mathrm{H_{\alpha}}$\,=\,0.160\,$\pm$\,0.005 (following \citet{2009A&A...495..959B}). Using regions defined by \citet{1991ApJS...76..383D} and calibrations from \citet{1984ApJ...279..763N} we measure the core emission of the Ca II H and K lines around 3933 \r{A} and 3968 \r{A} to be log\,$R^{\prime}_{HK}$\,=\,-4.503\,$\pm$\,0.044. This value is consistent with a young active star \citep{2008ApJ...687.1264M}.\par
Finally, legacy spectra from the GALAH survey \citep{2021MNRAS.506..150B} found \teff\,=\,6034\,$\pm$\,77 K, log\,$g$\,=\,4.36\,$\pm$\,0.18, \met\,=\,0.08\,$\pm$\,0.06 and \vsini\,=\,15.6\,$\pm$\,2.2\,\kms.

\subsection{Gemini-South and SOAR: High resolution direct Imaging} \label{subsec:direct-imaging}

\begin{figure}[]
    \centering
    \includegraphics[width=\linewidth]{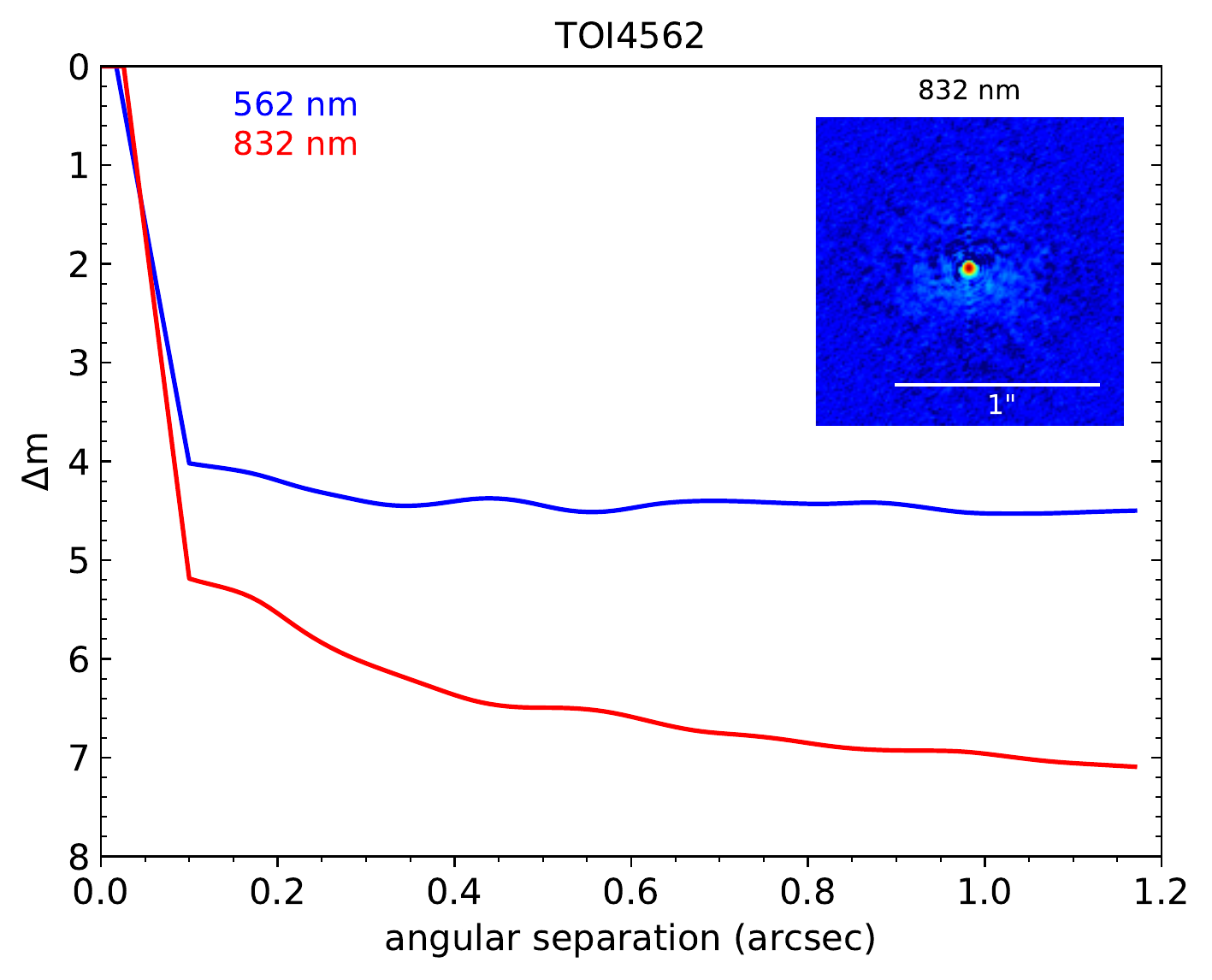}
    \includegraphics[width=\linewidth]{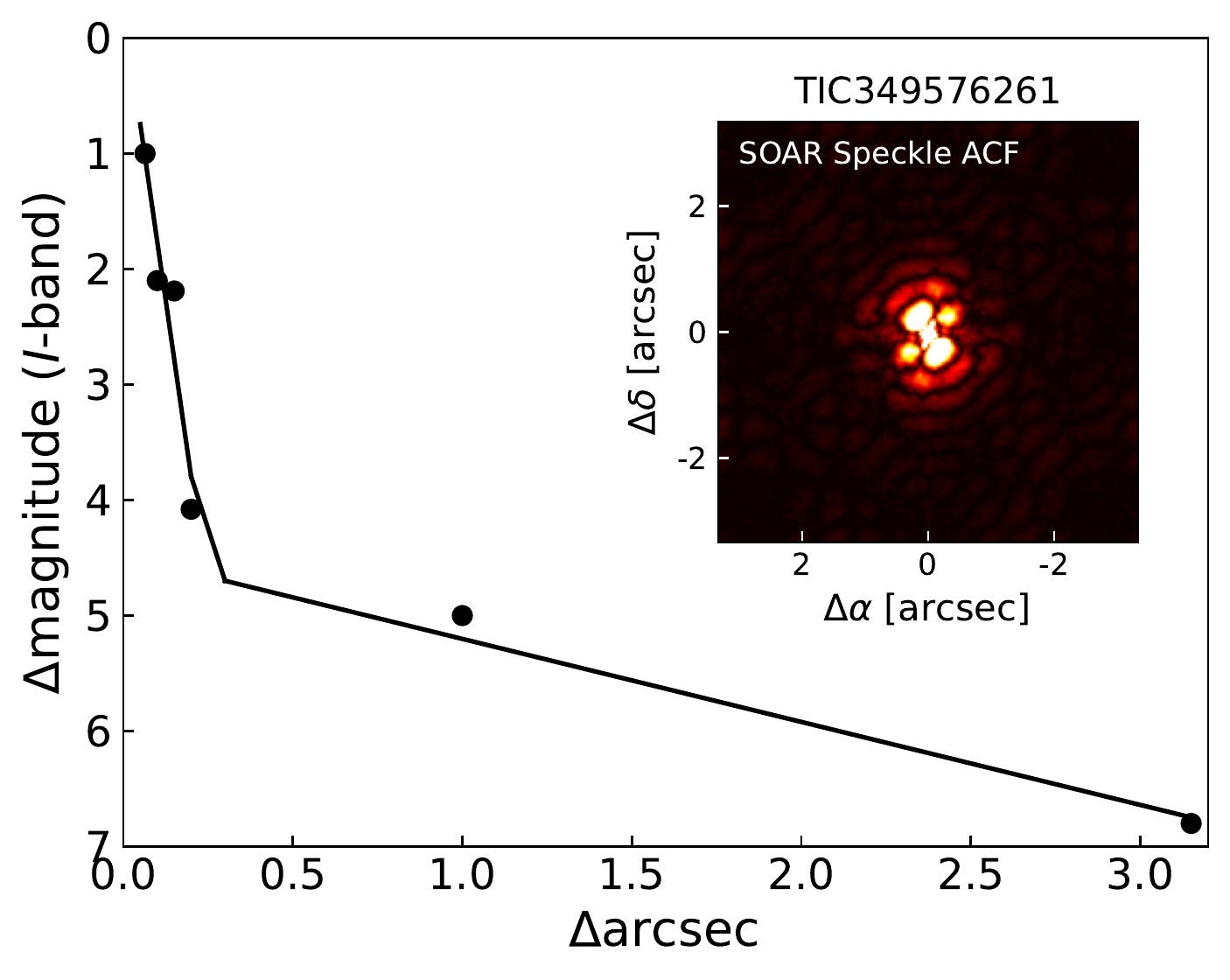}
    \caption{\textbf{Top} High resolution image of \toi obtained with the Zorro camera attached to the 8.1 m Gemini South telescope. The blue and red curves show difference in magnitude as a function of orbital separation from \toi obtained at wavelengths of respectively 562 and 832 nm. The inset plot shows the reconstructed image at 832 nm where no companion is detected. \textbf{Bottom} SOAR HRCam high resolution imaging of \toi. The difference in magnitude as a function of orbital separation from \toi is shown by the black line and the autocorrelation function on the inset image. There is no sign of a stellar sized companion to \toi. }\label{fig:highres}
\end{figure}

A first high resolution image of \toi was obtained on 2022-03-17 with the Zorro Speckle camera on the 8.1 m Gemini-South
telescope~\citep{2022FrASS...9.1163H} and is shown in the top of Figure~\ref{fig:highres}. Simultaneous observations were obtained at 562 and 832 nm respectively. Contrast curves were retrieved following \cite{2011AJ....142...19H} for both wavelengths and neither shows sign of a companion in the vicinity of \toib. A difference in magnitude $\Delta$m of 5 is achieved at a separation of $\sim0.1$". This allows us to rule out the presence of bright stellar objects in the same TESS pixel as \toi that would meaningfully impact the transit light curve to a projected distance of $\sim$ \sepvalue~au (given \toi's distance of \distvalue~pc). 

On 2022-04-19, another high resolution image was acquired with the HRCam instrument on the 4.1 m Southern Astrophysical Research (SOAR) telescope. \toi was observed as part of the SOAR TESS survey~\citep{2020AJ....159...19Z,2021AJ....162..192Z}, and the data was reduced following~\cite{2018PASP..130c5002T}. The image shown in the bottom panel of Figure~\ref{fig:highres} and shows a contrast in the $I$-band of 5 mag within 1" with no sign of a companion, in agreement with the Gemini-South observation.

\section{Age of \toi} \label{subsec:age}


\toi does not appear in the extensive list of stars with known age and/or belonging to associations and moving groups compiled from the literature in \cite{Bouma_2022}. Similiarly, we do not identify a co-eval population when applying the \textsc{comove} package~\citep{2021AJ....161..171T} that uses Gaia DR3 astrometric parameters to find whether a given possible young star candidate is co-moving with its visual neighbours.

This lack of evidence of \toi belonging to any known moving group or open cluster means its age estimation is challenging. The variability seen in both photometry and radial velocity are indicative of the presence of rotationally modulated surface brightness features, likely due to the presence of dark spots and bright plages/faculae. Combined with a fast rotation period (\Prot~= \Protvalue~days), this strongly suggests that \toi is a young and active star. 

Determining the age of a field star is notoriously difficult \citep{2010ARA&A..48..581S}. In the following paragraphs, we make use of the rotation and lithium abundance of \toi to qualitatively assess its youth. We note that though \toi exhibits signatures of activity and youth indicative of being younger than 1 Gyr, pin-pointing its age will remain difficult without placing it within co-moving populations. With increasingly more sophisticated clustering with updated \emph{Gaia} datasets, we hope that kinematics studies such as \citet{2017AJ....153..257O}, \citet{2018ApJ...856...23G}, \citet{2019AJ....158..122K} and \citet{2020AJ....159..166U} can provide improved census of young associations and groups. 

\begin{figure*}
    \centering
    \includegraphics{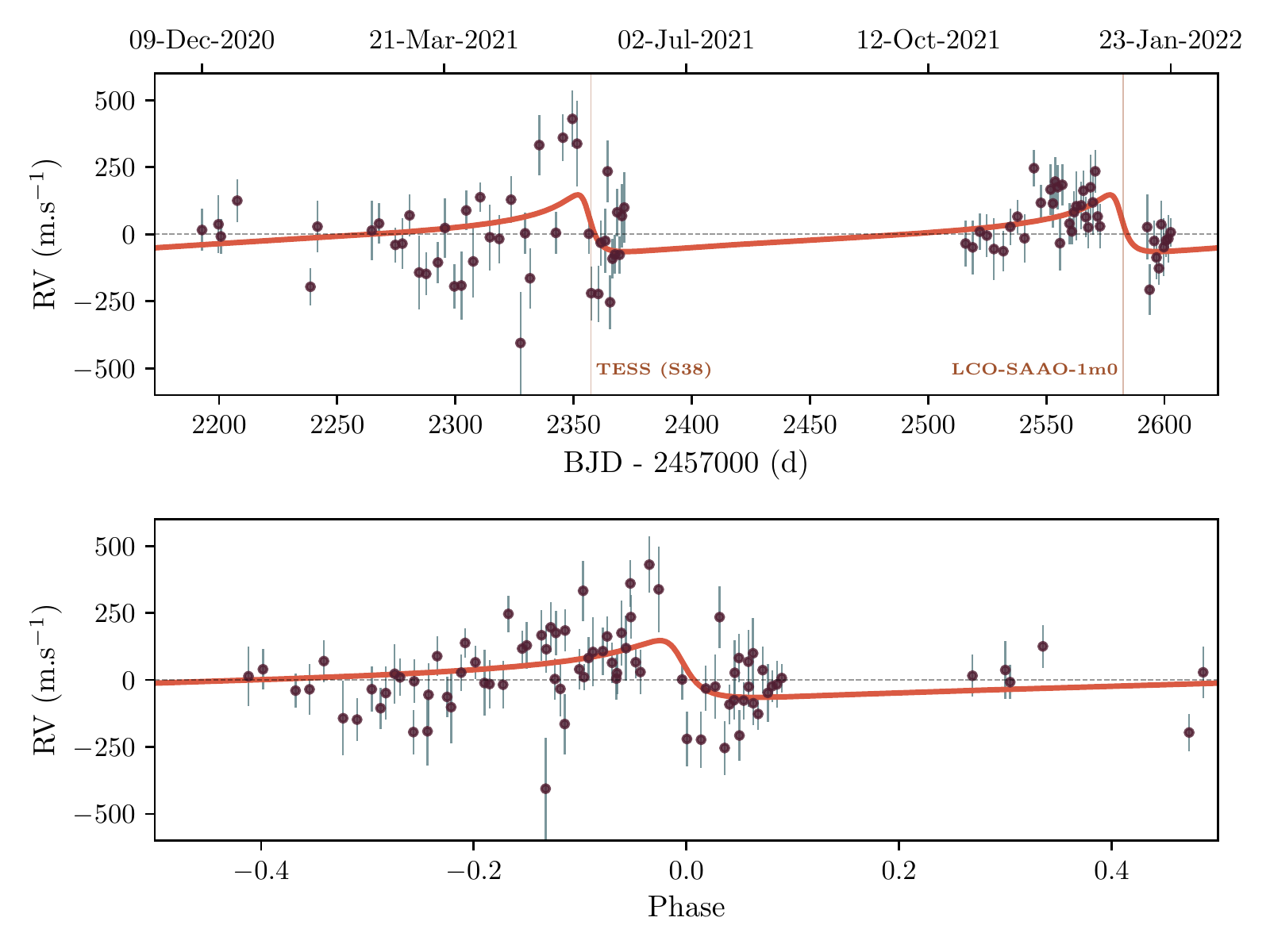}
    \caption{\textbf{Top}: 2-year radial velocity time series of \toi obtained with the CHIRON spectrograph (brown points) with the associated error bars. The Keplerian orbit fit from our global modelling is shown in red. Transits are highlighted in light brown. \textbf{Bottom}: Phase folded RVs (brown) with the Keplerian orbit best fits in red.}\label{fig:RV}
\end{figure*}

\begin{figure*}
    \centering
    \subfigure{\includegraphics[width=0.8\linewidth]{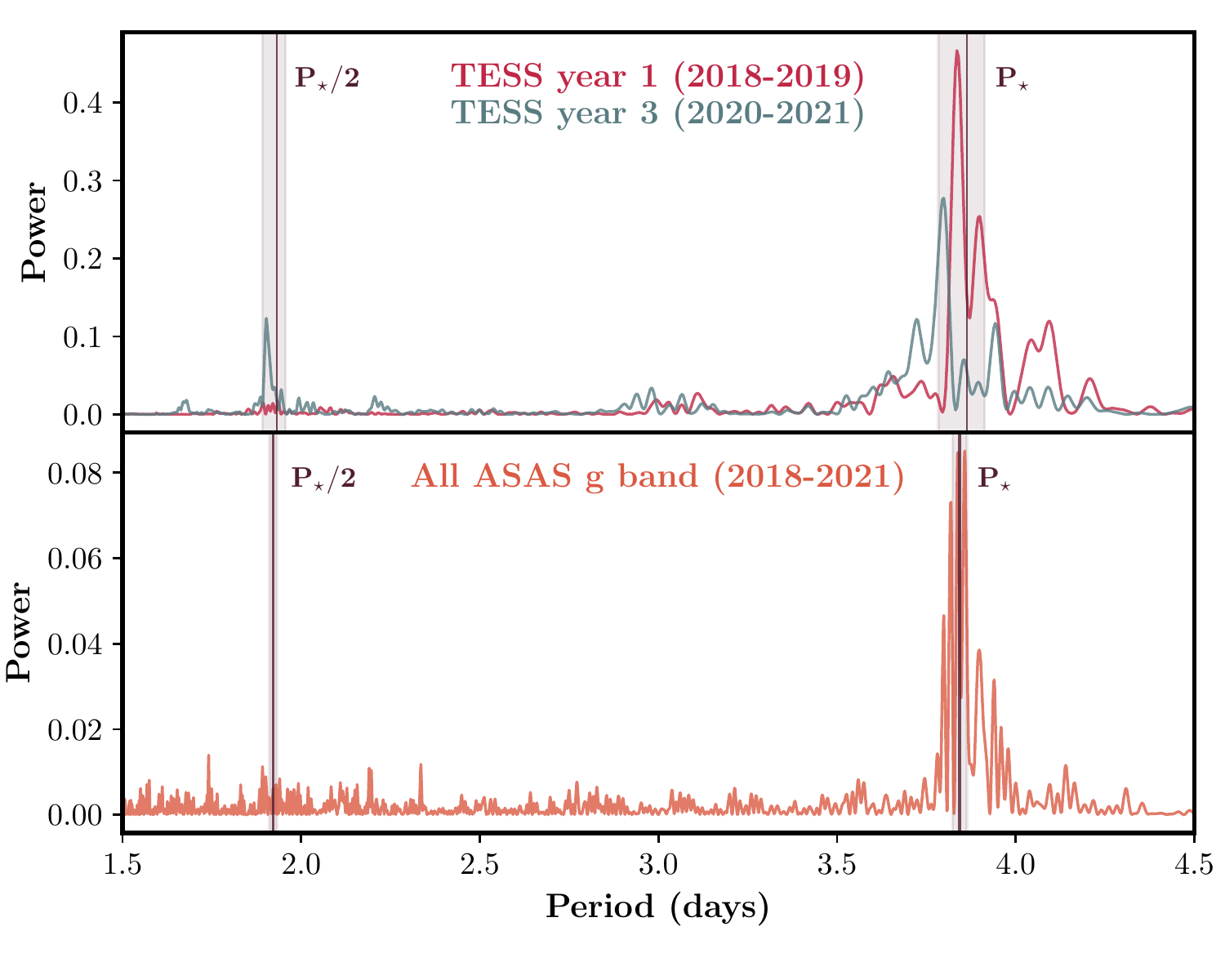}} 
    \subfigure{\includegraphics[width=0.5\linewidth]{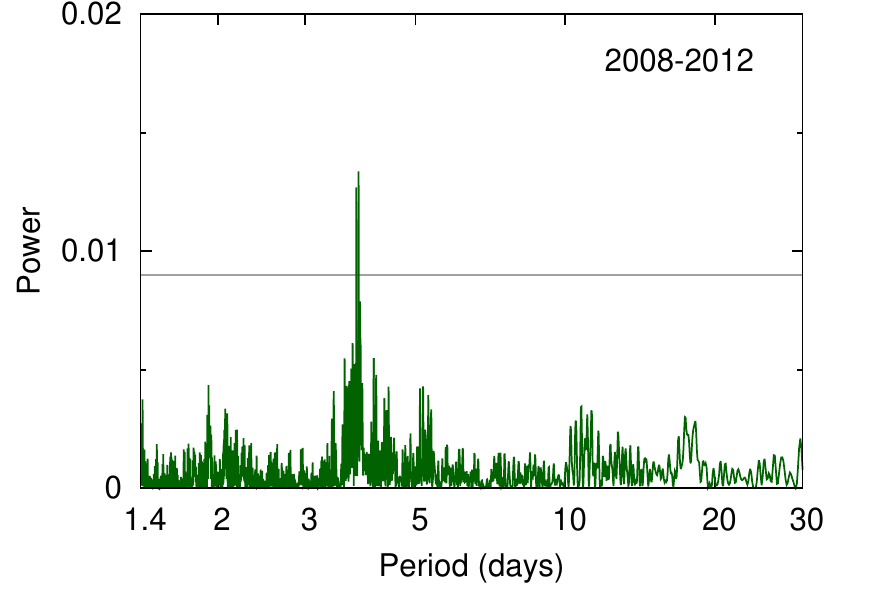}}
    \caption{\textbf{Top} GLS periodogram of \toi's photometry from \TESS for the first (2018-2019, red line) and third (2020-2021, blue line) year of data. The identified stellar rotation period (\Prot~= \Protvalue) and the first harmonic (\Prot$/2$) are shown with the vertical brown lines with the surrounding shaded areas  representing uncertainties. 
    \textbf{Middle} GLS periodogram of ASAS-SN's photometry gathered between 2018 and 2021 (orange line). The identified stellar rotation period (\Prot~= \ProtvalueASAS) and the first harmonic (\Prot$/2$) are shown as the vertical brown lines with their respective uncertainties (shaded area). 
    \textbf{Bottom} Periodogram on the 4 years (2008-2012) of WASP data yielding \Prot~estimates between 3.74 and 3.84 depending on the considered year. \\
    In all three periodograms, we notice a strong peak multiplicity around 3.9 days that could be associated with stellar surface features tracing differential rotation.}
    \label{fig:per_TESS}
\end{figure*}

\begin{deluxetable*}{llccc}
\tablenum{1}
\tablecaption{\toi parameters.}
\label{tab:star}
\tablewidth{0pt}
\tablehead{
\colhead{Parameters} & \colhead{Description} & \colhead{Prior} & \colhead{Value}
 & \colhead{Reference}
}
\decimals
\startdata
\textbf{Name and position} & & \\
TOI & \TESS Object of Interest & - &  4562 & \\
TIC & \TESS Input Catalog & - &  349576261 & ST18 \\
$Gaia$ & DR2 Source ID & - & 5288681857665822080 & \emph{Gaia} EDR3\\ 
RA & Right ascension (HH:MM:SS, J2000, epoch 2015.5) & - & 07:28:02.41 & \emph{Gaia} EDR3\\ 
DEC & Declination (DD:MM:SS, J2000, epoch 2015.5) & - & -63:31:04 & \emph{Gaia} EDR3\\ 
$\mu_{RA}$ & RA proper motion (\masyr) & - & -5.899\,$\pm$\,0.015 & \emph{Gaia} EDR3\\ 
$\mu_{DEC}$ & DEC proper motion (\masyr)  & - & 10.491 $\pm$ 0.011 & \emph{Gaia} EDR3\\
Type & Spectral type  & - & F7V  & \\
$\varpi$ & Parallax (mas) & $\mathcal{G}[2.857,0.05]$\tablenotemark{a} & \parallaxvalue & This work\\
D & Distance (parsec) & - & \distvalue & This work\\
\textbf{Photospheric parameters} & & \\
\teff & Effective temperature (K) & - & $6096 \pm 32$\,K & This work (CHIRON) \\
log $g$ & Surface gravity (dex) & - & \loggvalue & This work \\
\met & Bulk metallicity (dex) & $\mathcal{G}[0.2,0.3]$ & \fehvalue & This work\\
\vsini & Rotational velocity (\kms) & - & \vsiniCHIRON & This work (CHIRON)\\
\textbf{Physical parameters} & & \\
\Mstar & Mass (\Msun) &  $\mathcal{U}[0,2]$ & \mStarvalue & This work\\
$R_\star$ & Radius (\Rsun) &  &\rStarvalue & This work\\
$Age$ & Age (Myr) &  & \agevalue & This work \\
\textbf{Activity parameters} & & \\
\Prot & Equatorial rotation period (days) & - & \Protvalue & This work\\
Li 6708 EW & Li doublet ($\sim$ 6708 \r{A}) equivalent width (\r{A}) & - & 0.076\,$\pm$\,0.022 & This work \\
log $R'_{HK}$ &  & - & -4.503 $_{-0.052}^{+0.028}$ & This work (FEROS)\\
\textbf{Photometric parameters} & & \\
E(B-V) &  Interstellar extinction (mag) & $\mathcal{U}[0,0.1542]$\tablenotemark{b} & \ebvvalue & This work \\
\Tmag &  \TESS \Tmag (mag) & - & 11.533 $\pm$ 0.006 & ST18\\
\Vmag &  Johnson \Vmag (mag) & - & 12.098 $\pm$ 0.014 & H16\\
\Bmag &  Johnson \Bmag~(mag) & - & 12.698 $\pm$ 0.025 & H16\\
\gaiaG & Gaia \gaiaG (mag) & - & 11.948 $\pm$ 0.020\tablenotemark{c} & \emph{Gaia} EDR3\\
\gaiaBP & Gaia \gaiaBP (mag) & - & 12.262 $\pm$ 0.020\tablenotemark{c} & \emph{Gaia} EDR3\\
\gaiaRp & Gaia \gaiaRp (mag) & - & 11.467 $\pm$ 0.020\tablenotemark{c} & \emph{Gaia} EDR3\\
\Jmag &  2MASS \Jmag (mag) & - & 10.931 $\pm$ 0.023 & SK06 \\
\Hmag &  2MASS \Hmag (mag) & - & 10.693 $\pm$ 0.025 & SK06 \\
\Kmag &  2MASS \Kmag (mag) & - & 10.619 $\pm$ 0.023 & SK06 \\
\Wi &  WISE \Wi (mag) & - &  10.578 $\pm$ 0.023 & W10,C13 \\
\Wii &  WISE \Wii (mag) & - &  10.618 $\pm$ 0.020 & W10,C13 \\
\Wiii &  WISE \Wiii (mag) & - &  10.590 $\pm$ 0.061 & W10,C13 \\
NUV & GALEX/NUV calibrated AB magnitude (mag) & - & 17.133 $\pm$ 0.023 & B17\\
\hline
\enddata
\tablenotetext{a}{Adopted from Gaia EDR3 \cite{Gaia2016a,Gaia2020a}, corrected using a zero point offset value of -0.024 mas, following}~\cite{Lindegren2021}
\tablenotetext{b}{Adopted from \cite{2011ApJ...737..103S}}
\tablenotetext{c}{These are inflated uncertainties from the \emph{Gaia} photometric bands, following the convention from \cite{2013PASP..125...83E}}
\tablenotetext{}{\textbf{Priors:} $\mathcal{U}$ [$a,b$] uniform priors with boundaries $a$ and $b$; $\mathcal{G}$[$\mu$,$\sigma$] Gaussian priors}
\tablenotetext{}{\textbf{References:} \emph{Gaia} EDR3 \cite{Gaia2016a,Gaia2020a}; ST18 \cite{2018AJ....156..102S}; SK06 \cite{2006AJ....131.1163S}; W10 \cite{2010AJ....140.1868W}; C13 \cite{2014yCat.2328....0C}; B17 \cite{2017ApJS..230...24B};  H16 \cite{2016yCat.2336....0H}}
\end{deluxetable*}

\subsection{Stellar rotation and Gyrochronology} \label{subsec:gyro}

Young stars on the zero-age main-sequence spin rapidly. Over the course of a few billion years, mass loss from stellar winds spin-down Sun-like stars. The rotation period of Sun-like stars can be a tracer for their age. Rotation--colour--age relationships such as those from \citet{2007ApJ...669.1167B} and \citet{2008ApJ...687.1264M} are calibrated against co-eval clusters and associations, and can provide useful metrics to estimate stellar ages. Recent theoretically-motivated models, which are based in wind braking models and can incorporate core-envelope coupling, also provide such relationships \citep[e.g.][]{2020A&A...636A..76S}.

The 25 sectors of observations gathered by \TESS provide the means for a good estimation of the rotation period of \toi. As shown in Figure~\ref{fig:per_TESS}, we ran the \textsc{PyAstronomy} implementation of the Generalized Lomb-Scargle periodogram (GLS, \citealt{Lomb1976,Scargle1982, 2009A&A...496..577Z}) on the entire dataset and measured a rotation period of \Prot~= \Protvalue~days. To obtain \Prot, we first computed separate GLS periodograms for individual sectors (each covering $\sim$ 7 \Prot) previously binned to 10 minute cadence. The median and 1-$\sigma$ values of all obtained periods were then used to derive \Prot~and the associated uncertainties.

We note the clear second periodogram peak on Figure~\ref{fig:per_TESS}, close to \Prot. This could be showing differential rotation (i.e., the variation of \Prot~as a function of stellar latitude). This has been largely observed in Kepler stars~\citep{2013A&A...560A...4R}. We could suppose that the rotational modulation of two distinct clumps of surface stellar spots evolving at a different latitude would be at origin of the double peak \citep{1993A&A...269..351L}. 

In addition, \toi received 4 years of monitoring with the Wide Angle Search for Planets (WASP) Consortium \citep{2006PASP..118.1407P} Southern SuperWASP facility from 2008-2012. WASP-South is located at SAAO, and consists of an array of eight commonly mounted 200\,mm f/1.8 Canon telephoto lenses, each with a $2\mathrm{K} \times 2\mathrm{K}$ detector. A period analysis of the WASP-South light curves reveals periods of 3.74, 3.84, 3.82 and 3.84 days for the 2008/2009, 2009/2010, 2010/2011 and 2011/2012 respectively, in agreement with the \TESS light curves. The long term stability of the signal helps to confirm it as the correct alias of the rotational modulation signal. We note that WASP did not cover any transit event. 

Finally, we run periodograms on the available light curves from the All-Sky Automated Survey for Supernovae (ASAS-SN, \cite{2014AAS...22323603S,2019MNRAS.485..961J}). Sloan g-band data spanning from October 2017 to April 2022 shows very strong peaks in the periodogram at 3.84, 3.87 and 3.82 days for the 2018/2019, 2019/2020 and 2020/2021 seasons respectively (2017 and 2022 datasets are of poorer quality), agreeing with the other photometric datasets. Johnson V-band data was obtained between October 2016 and September 2018. Despite being less extensive and less densely sampled than the g-band photometry, a moderate peak (FAP $\sim$\,0.2\%) is found at 3.64 days, close to \Prot.

Using the age-rotation relationship from \cite{2008ApJ...687.1264M}, we found \toib to be 110-490 (3$\sigma$) Myr old. We note that age estimates from this relationship assumes that the star lies on the slow-sequence of the age-rotation relationship. Stars are often found to be more rapidly rotating than such sequences for a given age, which has been attributed to binarity in cluster populations  \citep[e.g.,][]{2016ApJ...822...47D,2020MNRAS.492.1008G}. Though there is no evidence for \toi being part of a binary system, caveats still apply for gyrochronology-based age estimates. For a 1.2 \Msun~star with \Prot = \Protvalue, the model from~\cite{2020A&A...636A..76S} gives a consistent age estimate of 300-400 Myr.

\begin{figure}[]
    \centering
    \includegraphics[width=\linewidth]{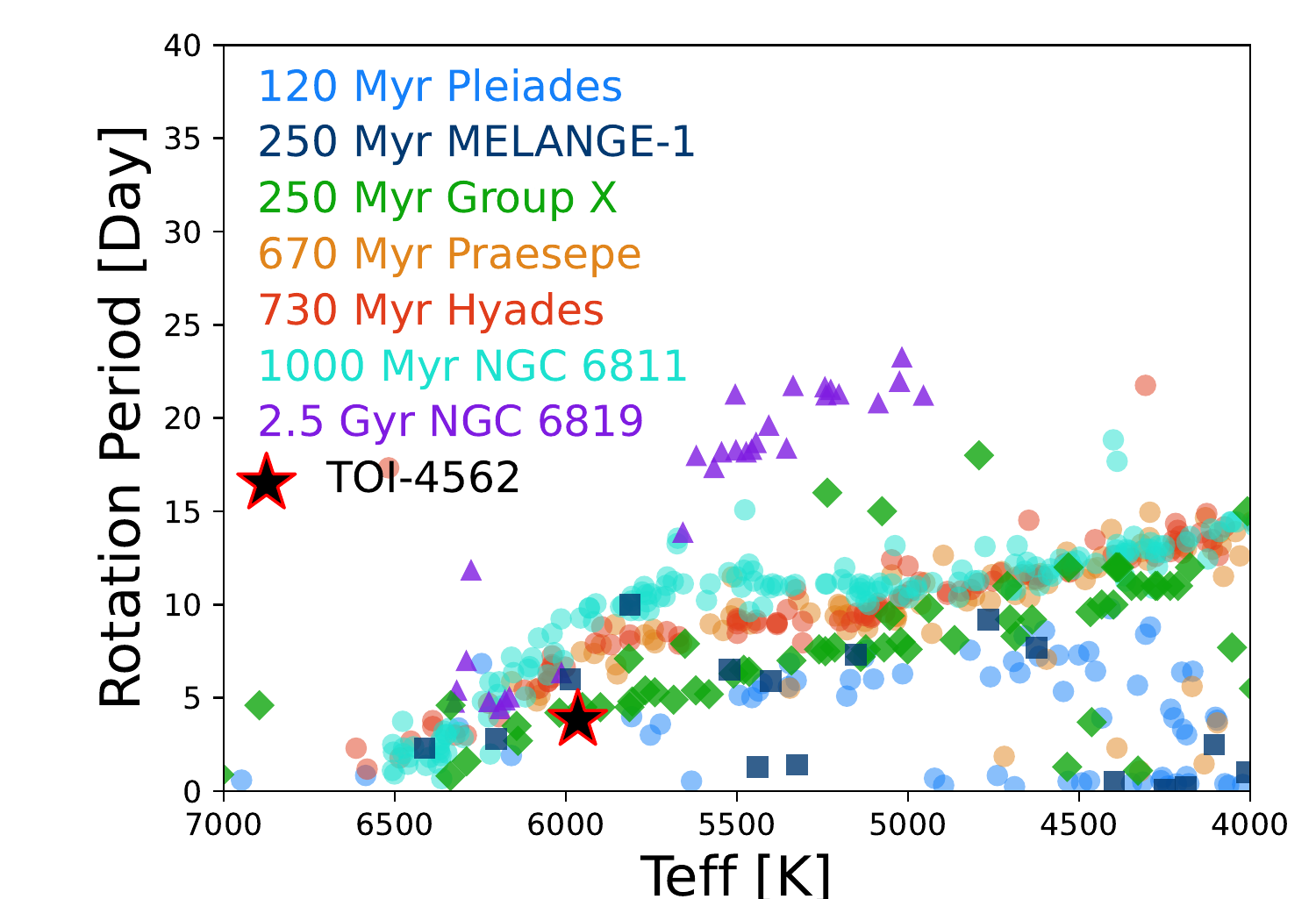}
    \includegraphics[width=\linewidth]{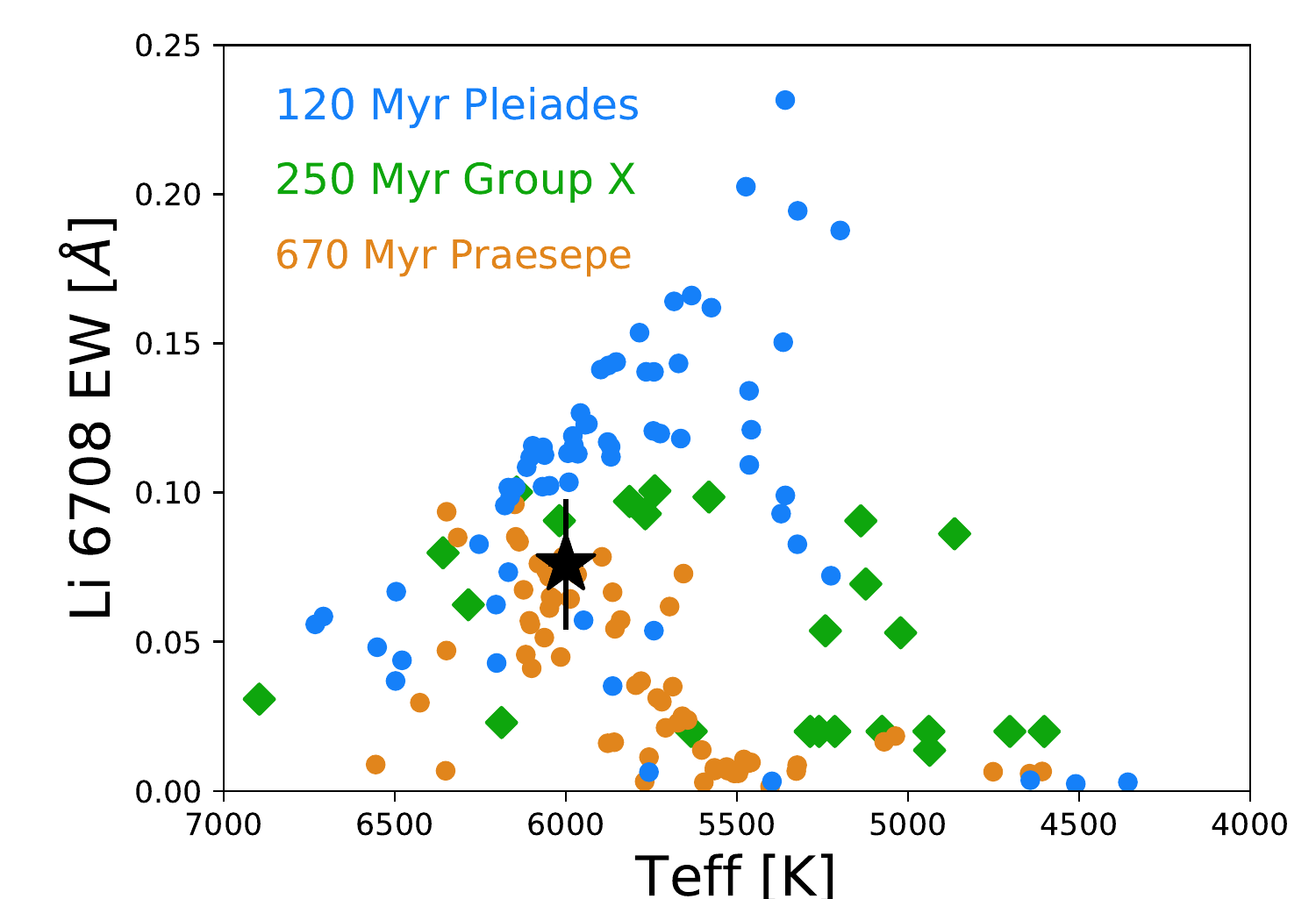}
    \caption{Youth indicators for TOI-4562. \textbf{Top:} The rotation period of TOI-4562 compared to the distribution of stars within known associations and clusters, including the Pleiades \citep{2016AJ....152..113R}, MELANGE-1 \citep{2021AJ....161..171T},  Group X \citep{2022arXiv220606254N,2022A&A...657L...3M}, Praesepe and Hyades \citep{2016ApJ...822...47D, 2019ApJ...879..100D}, NGC 6811 \citep{2019ApJ...879...49C} and NGC 6819 \citep{2015Natur.517..589M}. \textbf{Bottom:} Equivalent width of the lithium doublet at 6707.76 and 6707.91 \r{A} for \toi (red star) and stars in the Praesepe (orange, \citealt{2017AJ....153..128C}), Group X~\citep{2022arXiv220606254N}, and Pleiades (black, \citealt{2018A&A...613A..63B}) clusters. \toi lies at an age comparable to the Hyades and Praesepe.}\label{fig:Liew}
\end{figure}

The top plot of Figure~\ref{fig:Liew} shows the rotation period of \toi compared with stars of known nearby clusters and associations. \toi's \Prot\,is consistent with that of stars belonging to Group X~\citep{2022arXiv220606254N,2022A&A...657L...3M}, with an estimated age of 300 Myr.

\begin{figure}
    \centering
    \includegraphics[width=1\linewidth]{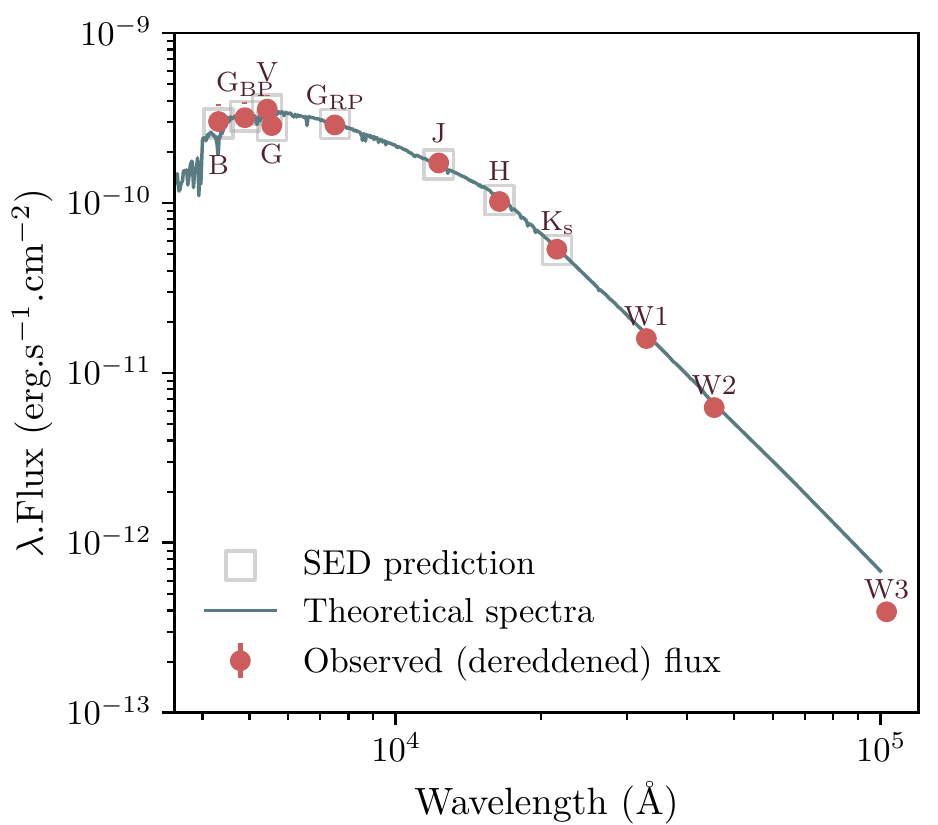}
    \caption{Spectral energy distribution (SED) of \toi b. Red points are the observed magnitudes in different wavelength bands (labelled with corresponding letters) corrected from interstellar reddening. Predicted magnitudes from the isochrone part of our global model are shown as grey squares. The blue line is a theoretical spectra for a star with \teff$= 6000$ K, log $g$ = 4.5~dex and [M/H] = 0; adopted from \cite{2014MNRAS.440.1027C}.}\label{fig:SED}
\end{figure}

\subsection{Lithium} \label{subsec:Li}
The convective envelope of low-mass stars (\Mstar\,$<$\,1.5 \Msun) allows efficient transport of lithium to deeper and hotter regions in a star's interior, where it gets destroyed by proton capture. Calibrated with stars in clusters and associations, this lithium depletion can be used as a proxy for stellar age. Using CHIRON spectra (see section \ref{subsec:chiron}), we measured the equivalent width of the lithium doublet at 6707.76 and 6707.91 \r{A}. We fit two Gaussian line profiles of the same depth at the respective wavelengths of the lithium doublet and one auxiliary with a different depth to account for the nearby Fe I line at 6707.43 \r{A}, usually blended with the Li doublet. All profiles share the same width as per the rotational broadening of the star. We measure a lithium equivalent width of 0.084 $\pm$ 0.007 \r{A} from a median combined spectrum of all our CHIRON observations. These data are displayed in Figure~\ref{fig:Liew}. 
On the same figure, we show the lithium equivalent width as function of effective temperature for stars belonging to clusters with well constrained ages, the Pleiades ($\sim$\,125 Myr), Group X ($\sim$\,300 Myr) and Praesepe ($\sim$\,670 Myr). \toi exhibits a Li equivalent width shallower than most of the Pleiades stars and of comparable strength to stars from the Praesepe cluster, at an effective temperature of 6000 K. Combined with the Gyrochronology analysis, \toi's age is consistent with a star younger than the Praesepe/Hyades clusters (i.e., $\lesssim$ 700 Myr).

\subsection{Lithium, log\,$R^{\prime}_{HK}$\ and $B - V$}
\label{subsec:Li+logrhk}
Finally, we used the more recent \textsc{baffles} package~\citep{2020ApJ...898...27S} to derive an age estimate. \textsc{baffles} is a Bayesian framework in which lithium abundance, the log\,$R^{\prime}_{HK}$ index and $B - V$ colour are used to infer an age posterior for the studied star. The likelihood functions used are calibrated against star belonging to open clusters and associations (i.e., with well constrained ages). For log\,$R^{\prime}_{HK}$\,=\,-4.503 (from the FEROS spectra), Li 6708 EW = 84\,$\pm$\,7 m\r{A} and $B - V$\,=\,0.52\,$\pm$\,0.03 (corrected from extinction), we recover a 1-$\sigma$ age estimate of 340-715-1500 Myr. This is slightly higher than our previous estimate, but we note that \textsc{baffles} uses an age posterior extending by default to 12 Gyr. Comparing to the star sample used  in~\cite{2020ApJ...893...67M}, belonging to well studied star clusters, \toi's photometric variability and rotation period are respectively larger and smaller than any stars of age 1 Gyr, therefore pointing at a younger age. Applying a conservative maximum age cut off of 1.5 Gyr to the age posterior when running 
\textsc{baffles} yields a 1-$\sigma$ age of 320-619-1100 Myr, closer to previously quoted values.

\section{Analysis and Results} \label{sec:analysis}

To best determine the system properties of \toi, we perform a joint modeling of all available photometric and spectroscopic datasets, including stellar isochrone models that constrain the properties of the host star. The paragraphs below detail individual components of this model.

\subsection{Transit modeling} \label{subsec:transits}

Despite the 225 day orbital period of \toib, the extensive observations of \toi by \TESS allowed four transits to be observed. Spot modulated variability at the $\sim$3\% level is seen on the \TESS light curve due to the active nature of \toi, as expected given its young age. For the purposes of the transit modeling, we detrend the region around each transit epoch with a fourth-order polynomial. The polynomial is fitted using the out-of-transit regions of the light curve within 0.5\,days of the transit center.
We model the transits as per \citet{2002ApJ...580L.171M} via the \textsc{batman} package \citep{2015PASP..127.1161K}. Free parameters that describe the transit model include the transit centre \tc\  at each transit epoch, radius ratio $R_{\mathrm{p}}$/$R_{\mathrm{\star}}$, line of sight inclination of the transit $i$, and the eccentricity parameters $\sqrt{e}\cos\omega$ and $\sqrt{e}\sin \omega$. A quadratic model was used to account for Limb Darkening using coefficients $\mu_{1TESS}$ and $\mu_{2TESS}$ fixed to those interpolated from \cite{2017AA...600A..30C} at the atmospheric parameters of \toi for the \TESS transits. We note that $a$/$R_{\mathrm{\star}}$ was not directly sampled but rather computed from the free parameters \Porb, \Mstar, \Rstar\,and planet mass \Mp. For the two (same epoch, different telescopes) SAAO LCOGT transits, the Limb Darkening coefficients $\mu_{1_{LCO}}$ and $\mu_{2_{LCO}}$ are computed for the SDSS i' band from \citet{2011AA...529A..75C}, using the interpolation routine from \cite{2013PASP..125...83E} with \teff\,=\,6000\,K, log $g$\,=\,4.5 and [M/H]\,=\,0.1, computed with the least square method (LSM). For the SAAO LCOGT data, we also incorporate the effects of instrumental systematic variations that are common to ground-based photometric observations via a simultaneous detrending of the light curve against parameters describing the observation airmass to which we add a linear trend with respect to time. All detrended light curves and the best transit model fits are shown in Figure~\ref{fig:transits}.

\subsection{Eccentric orbit validation} \label{subsec:ecc}

Even though the transit shape points to a very eccentric orbit, this can be degenerate with the mean stellar density \rhostar. To cross validate the eccentric nature of \toib's orbit, we compared the stellar density obtained from isochrone fitting, which we call \rhostar, to the stellar density obtained from a forced circular orbit, that we call \rhocirc. To obtain \rhocirc, we run a transit only model, imposing $e$ = 0 and with the following free parameters: \rhocirc, transit center \tc~for each transits, radius ratio \RprStar, impact parameter $b$, limb darkening coefficients ($\mu_{1_\mathrm{LCO}}$, $\mu_{2_\mathrm{LCO}}$, $\mu_{1_\mathrm{TESS}}$ and $\mu_{2_\mathrm{TESS}}$) and instrument systematics parameters (for the SAAO LCOGT transits). Leveraging this photo-eccentric effect~\citep{2012ApJ...761..163D} allow us to recover the space of possible parameters for $e$ and $w$ (see \citealt{2015ApJ...798...66D}). On~Figure~\ref{fig:corner_e_w}, we show in blue the resulting 2d distribution of $e$ and $w$ that agree with the derived ratio for~\rhocirc/\rhostar, given\footnote{In the case of \toib, the condition $(\frac{a}{R_{\star}})^2 \gg \frac{2}{3} \left( \frac{1+e}{1-e} \right)^3 $, required for~\ref{eq:asteroprofiling} to be valid (see~\citealt{2014MNRAS.440.2164K}) is fulfilled.}:
\begin{equation}
\label{eq:asteroprofiling}
\frac{\rho_{\mathrm{circ}}}{\rho_{\star}} = \frac{( 1 + e~\sin w)^3}{( 1-e^2 )^{3/2}}
\end{equation}
This grants us increased confidence of the true eccentric nature of \toib's orbit.

\subsection{Radial velocity modeling} \label{subsec:rv}

The radial velocities obtained over the 2 consecutive orbits of \toib were modeled using a Keplerian orbit. Some fitted parameters are shared with the transits and stellar isochrone fitting, such as \tc, \Porb, \ars (derived), \Rstar, \Mstar, $i$, \secosw~and \sesinw. To model the velocities, we add the planet mass \Mp, a radial velocity offset \offsetchiron, and a white noise term, $\sigma_{Y1}$. The semi-amplitude of the planetary signature \Kamp~was computed from the above parameters. The orbital solution and the associated likelihood from the fit to the data are computed from \Kamp, \tc, \Porb, \secosw~and \sesinw~via the \textsc{radvel} package \citep{2018PASP..130d4504F}.\par
We also try to add a Gaussian Process using a Quasi-Periodic kernel, implemented through \textsc{radvel} to model the stellar noise apparent in the data. The resulting parameter values do not yield a significant difference, therefore not justifying the necessity to use a correlated noise model to account for the stellar intrinsic variability seen in the radial velocities. With one datapoint a day at most, the sampling is too sparse for the Gaussian Process to correctly grasp the $\sim$ 4 days stellar period. Crudely assuming a spot covering 0.6-1.2\% ($\delta_{spot}$) of the stellar surface, we can approximate an activity induced radial velocity semi-amplitude $K_{act}$ of $v$ sin $i\times \delta_{spot} \sim 100-200$ \ms, comparable to the jitter level seen in Figure~\ref{fig:RV}. 

We attempted to fit a second longer period circular planet to the radial velocities. We used uniform priors for the period ($\mathcal{U}[300:2000]$ days), planet mass ($\mathcal{U} [0.002:0.1]$~\Msun) and $t_0$ ($\mathcal{U}[1398:3398]$ TBJD). The posterior distribution are not clearly converging, favouring larger periods and smaller masses. With a \Kamp~of $\sim$ 70 \ms, the best solution is clearly below the activity level and therefore not trustworthy. Long term data is needed to attempt to constrain a longer period companion.

\subsection{Spectral energy distribution model} \label{subsec:sed}
To constrain the host star parameters \Rstar, \Mstar, \met~and \teff\,we also model the spectral energy distribution of \toi simultaneously to the transit and radial velocity models. The stellar parameters are modeled using the MESA Isochrones \& Stellar Tracks \citep{2011ApJS..192....3P,2013ApJS..208....4P,2015ApJS..220...15P,2016ApJ...823..102C}. We interpolate evolution tracks using the \textsc{minimint} package \citep{sergey_koposov_2021_5610692} against \Mstar, age, \met~and the photometric bands $B$, $V$, \emph{Gaia} $G$, $Bp$, $Rp$, 2MASS bands $J$, $H$, and $K$. \Rstar\,is derived from the isochrone predicted values for log\,$g$ and \Mstar. To account for uncertainties in the stellar evolution models, we adopt a 4\% uncertainty floor in stellar radius, and 5\% floor in stellar mass, where appropriate \citep{2022ApJ...927...31T}.  For the effective temperature \teff, we apply a Gaussian prior such that the predicted \teff\,interpolated from the isochrone is compared against that measured from the CHIRON spectra as an additional likelihood term. Predicted fluxes from the SED model are corrected for interstellar reddening with the \textsc{PyAstronomy} \textsc{unred} package, that uses the parameterization from \cite{1999PASP..111...63F}. Extinction is a free parameter, with a maximum value of $E(B-V) =$\,0.1542 mag, as estimated from the \citet{2011ApJ...737..103S} maps over a 5 arcmin radius\footnote{Obtained from the \href{https://irsa.ipac.caltech.edu/applications/DUST/}{NASA/IPAC Infrared Science Archive}} around \toi. We also incorporate a Gaussian prior on the distance modulus via the observed \emph{Gaia} parallax to \toi. We offset Gaia DR3's parallax value by -0.023861 mas, the parallax zero-point offset estimated using the routine from~\cite{Lindegren2021}\footnote{\url{https://gitlab.com/icc-ub/public/gaiadr3_zeropoint}} and function of ecliptic latitude, magnitude and colour. At each MCMC jump step, the observed spectral energy distribution is compared against the interpolated MIST model predictions for a given tested stellar parameter.

\subsection{Global model} \label{subsec:globalfit}
The global model includes simultaneous fits of the \TESS and ground-based photometric datasets (\ref{subsec:transits}), the CHIRON velocities (\ref{subsec:rv}), and stellar isochrone model (\ref{subsec:sed}), as shown in Figure~\ref{fig:transits},~\ref{fig:RV} and ~\ref{fig:SED} respectively. We explore the best fit parameters and the posterior distribution via the Affine Invariant Markov chain Monte Carlo Ensemble sampler \textsc{emcee} \citep{2013PASP..125..306F}. The resulting parameters for \toib are given in Table~\ref{tab:planet}. \par
\begin{figure}
    \centering
    \includegraphics[width=1\linewidth]{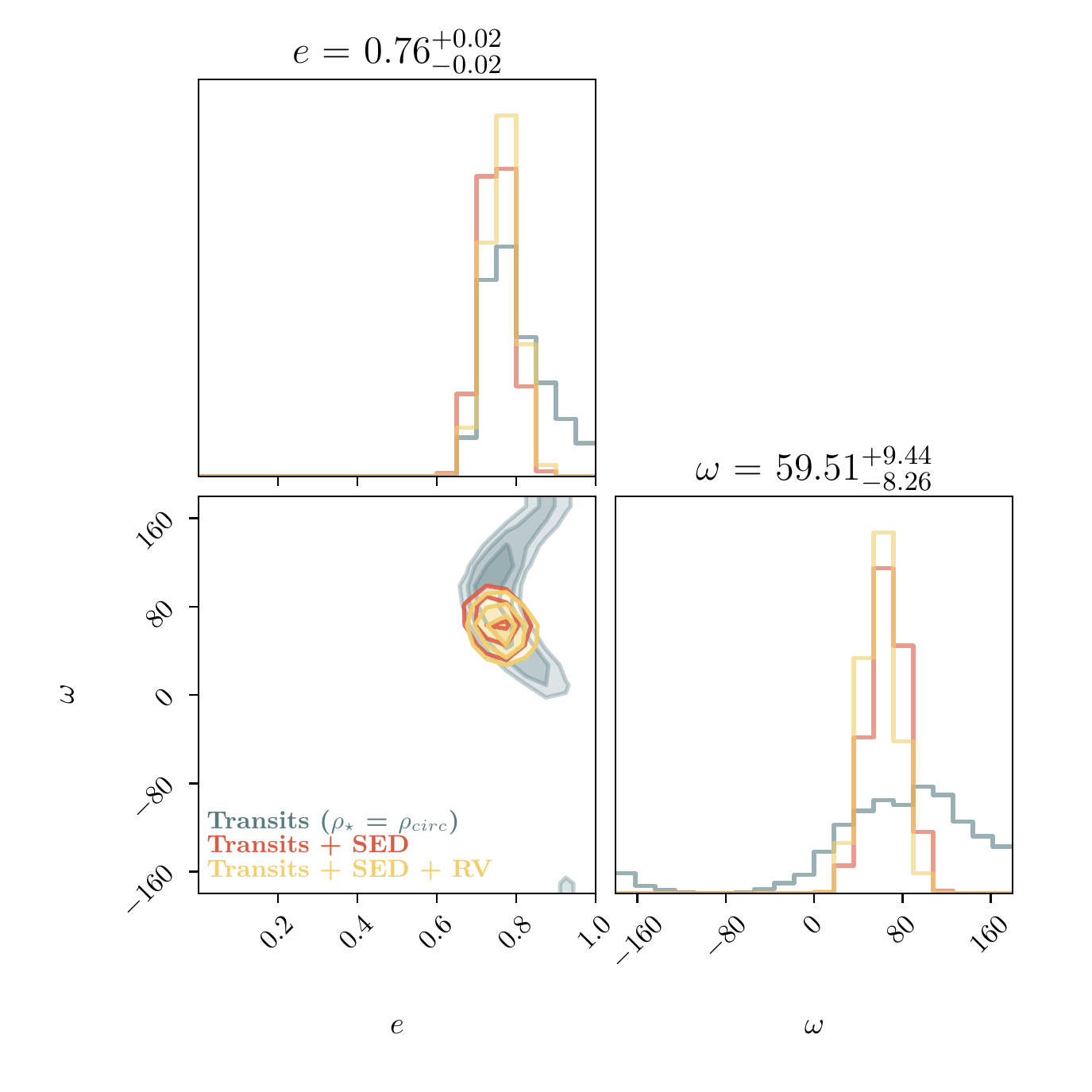}
    \caption{Posterior distributions for \toi b's orbital eccentricity ($e$) and argument at periapse ($\omega$). The blue distribution was obtained from comparing \rhostar~obtained from a SED fit and \rhocirc~resulting from a fit of only the photometric data (\TESS+ LCO-SAAO) with an imposed circular orbit. The red and yellow posterior distributions respectively results from a photometric data + SED fit and the complete model including the radial velocities.} \label{fig:corner_e_w}
\end{figure}

Figure~\ref{fig:corner_e_w} illustrates the contribution of each part of the model to constrain the high eccentricity of \toib. In blue we show the distribution stemming from the value of \rhostar~when imposing a circular orbit to the planet and fitting only for transits (see Section~\ref{subsec:ecc}). Then, a model containing both the transits and isochrones, shown in red, greatly constrain the eccentricity. Finally, the addition of RVs (shown in yellow) only slightly improves the constrain the eccentricity, due to both the large stellar activity and the poor coverage of the periastron passages.


\begin{deluxetable*}{llcc}
\tablenum{2}
\tablecaption{\toib parameters. \label{tab:planet}}
\tablewidth{0pt}
\tablehead{
\colhead{Parameters} & \colhead{Description} & \colhead{Priors} & \colhead{Values}
}
\decimals
\startdata
\textbf{Transit parameters} & & &\\
$T_{c,1}$ \tablenotemark{a} & Transit mid-time (\TBJD) &  $\mathcal{U}$[1456.83,1456.93] & \tcfirstvalue \\
$T_{c,2}$ \tablenotemark{a} & Transit mid-time (\TBJD) &  $\mathcal{U}$[1681.94,1682.04] & \tcsecondvalue \\
$T_{c,3}$ \tablenotemark{a} & Transit mid-time (\TBJD) &  $\mathcal{U}$[2132.17,2132.27] & \tcthirdvalue \\
$T_{c,4}$ \tablenotemark{a} & Transit mid-time (\TBJD) &  $\mathcal{U}$[2357.29,2357.39] & \tcfourthvalue \\
$T_{c,5}$ \tablenotemark{a} & Transit mid-time (\TBJD) &  
$\mathcal{U}$[2582.41,2582.51] & \tcfifthvalue \\
$T_c$ \tablenotemark{a} & Derived linear ephemeris &  $\mathcal{U}$[1456.83,1456.93] & \tcvalue\\
$P_{\mathrm{orb}}$ \tablenotemark{a} & Orbital period (days) & Derived linear ephemeris & \pervalue\\
$T_{14}$ & Transit total duration (hours) & - & \tdurvalue\\
$R_\mathrm{p} / R_\star$ & Radius ratio & $\mathcal{U}[0,0.2]$ & \rprstarvalue \\
$a/ R_\star$ & Normalised Semi-major axis, derived from [\Mstar,\Rstar,\Porb,\Mp]  & - & \smarStarvalue\\
$b$ & Impact parameter & - & \bvalue\\
$\delta$ & Transit depth (ppm) & - & \tdepthvalue\\
\secosw \tablenotemark{a} & Reparameterization of $e$ and $\omega$ & $\mathcal{U}[-1,1]$ & \secoswvalue\\
\sesinw \tablenotemark{a} & Reparameterization of $e$ and $\omega$ & $\mathcal{U}[-1,1]$ & \sesinwvalue\\
$i$ & Planet orbit inclination (\degree) & $\mathcal{U}[84,90]$ & \incvalue\\
$\mu_{1_{TESS}}$ \tablenotemark{b} & Quadratic Limb Darkening law coefficient 1 (TESS) & Fixed & 0.28 \\
$\mu_{2_{TESS}}$ \tablenotemark{b} & Quadratic Limb Darkening law coefficient 2 (TESS) & Fixed & 0.29 \\
$\mu_{1_{LCO}}$ \tablenotemark{c} & Quadratic Limb Darkening law coefficient 1 (LCO) & Fixed & 0.28 \\
$\mu_{2_{LCO}}$ \tablenotemark{c} & Quadratic Limb Darkening law coefficient 2 (LCO) & Fixed & 0.29 \\
\textbf{Radial velocities parameters} & & &\\
$K$ & RV Semi-amplitude (\ms) &  - & \kampvalue\\
$M_\mathrm{p}$ & Mass (\Msun) & $\mathcal{U}[0,0.2]$ & \mpvalue\\
$\gamma_{CHIRON}$ & RV offset (\ms) & $\mathcal{U}[5200,5400]$ & \meanvalue\\
$\sigma_{Y1}$ & RV jitter, first orbit (\ms) & $\mathcal{U}[0,600]$ &  \jitvalue\\
\textbf{Derived parameters} & & &\\
$R_\mathrm{p}$ & Radius (\REarth) & - & \rpearthvalue\\
 & Radius (\RJup) & - & \rpjupvalue\\
$M_\mathrm{p}$ & Mass (\MEarth) & - & \mpearthvalue\\
 & Mass (\MJup) & - & \mpjupvalue\\
$e$ & Eccentricity & - & \eccvalue\\
$w$ & Argument at periapse (\degree) & - & \wvalue\\
$\rho_p$ & Density (g~cm$^{-3}$) & - & \densityvalue\\
$a$ & Semi-major axis (AU) & - & \smaAUvalue\\
$\langle T_{\mathrm{eq}} \rangle$ & Temporal average equilibrium temperature (K) \tablenotemark{d} & - & \teqvalue\\
$T_{\mathrm{peri}}$ & Equilibrium temperature at periapsis (K) & - & \tperivalue\\
$T_{\mathrm{apo}}$ & Equilibrium temperature at apoapsis (K) & - & \tapovalue\\
\hline
\enddata
\tablenotetext{}{\textbf{Priors:} $\mathcal{U}$ [$a,b$] uniform priors with boundaries $a$ and $b$}
\tablenotetext{a}{Parameters common to the Transit and RV models}
\tablenotetext{b}{Adopted at the \TESS band from \citet{2017AA...600A..30C}, using ATLAS model with \teff~= 6000K, log $g$~= 4.5 and \met~= 0.1 and computed with the least square method (LSM)}
\tablenotetext{c}{Computed for the SDSS i' band from \citet{2011AA...529A..75C}, using the interpolation routine from \cite{2013PASP..125...83E} with \teff~= 6000K, log $g$~= 4.5 and \met~= 0.1 and computed with the least square method (LSM)}
\tablenotetext{d}{Computed for an elliptical orbit from \cite{2017ApJ...837L...1M}, using an albedo of $A$ = 0.4,  $\epsilon$ = 1 and $\beta$ = 0.74.}
\end{deluxetable*}

\begin{figure}
    \centering
    \includegraphics[width=1\linewidth]{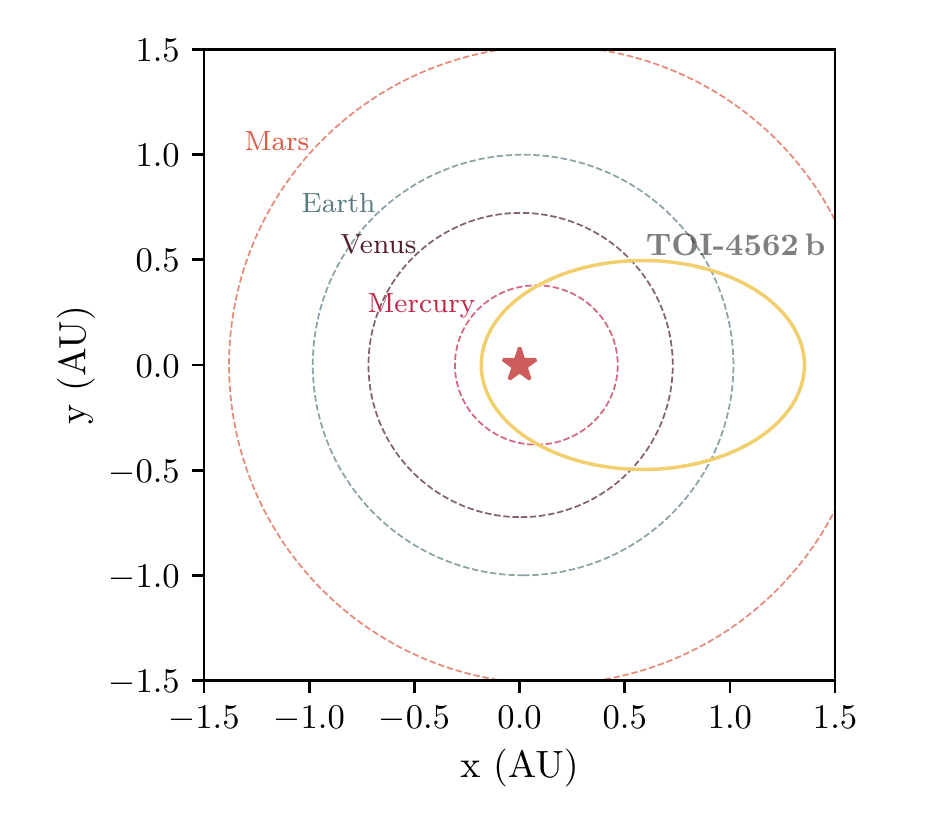}
    \caption{Two dimensional orbit of \toib (yellow solid line) compared with Mercury (red dashed line), Venus (purple dashed line), Earth (blue dashed line) and Mars (orange dashed line).}\label{fig:orbits}
\end{figure}

\section{Discussions and Conclusions} \label{sec:conclusion}

We report the discovery of \toib, a temperate gas giant on a highly eccentric orbit around a young Sun-like star. The planet has a mass of~\mpjupvalue~\MJup~and a radius of \rpjupvalue~\RJup. With an orbital period of \pervalue~days, it is to date the second longest period planet in the \TESS sample (after TOI-2180b, \citealt{2022AJ....163...61D}). \toib' resides in a highly elliptic orbit ($e$\,=\eccvalue), and has, based on~\citep{2020A&A...636A..76S}, an age younger than the Praesepe and Hyades clusters. A representation of its orbit alongside the inner Solar System planets is shown in Figure~\ref{fig:orbits}.

\begin{figure}
    \centering
    \includegraphics[width=1\linewidth]{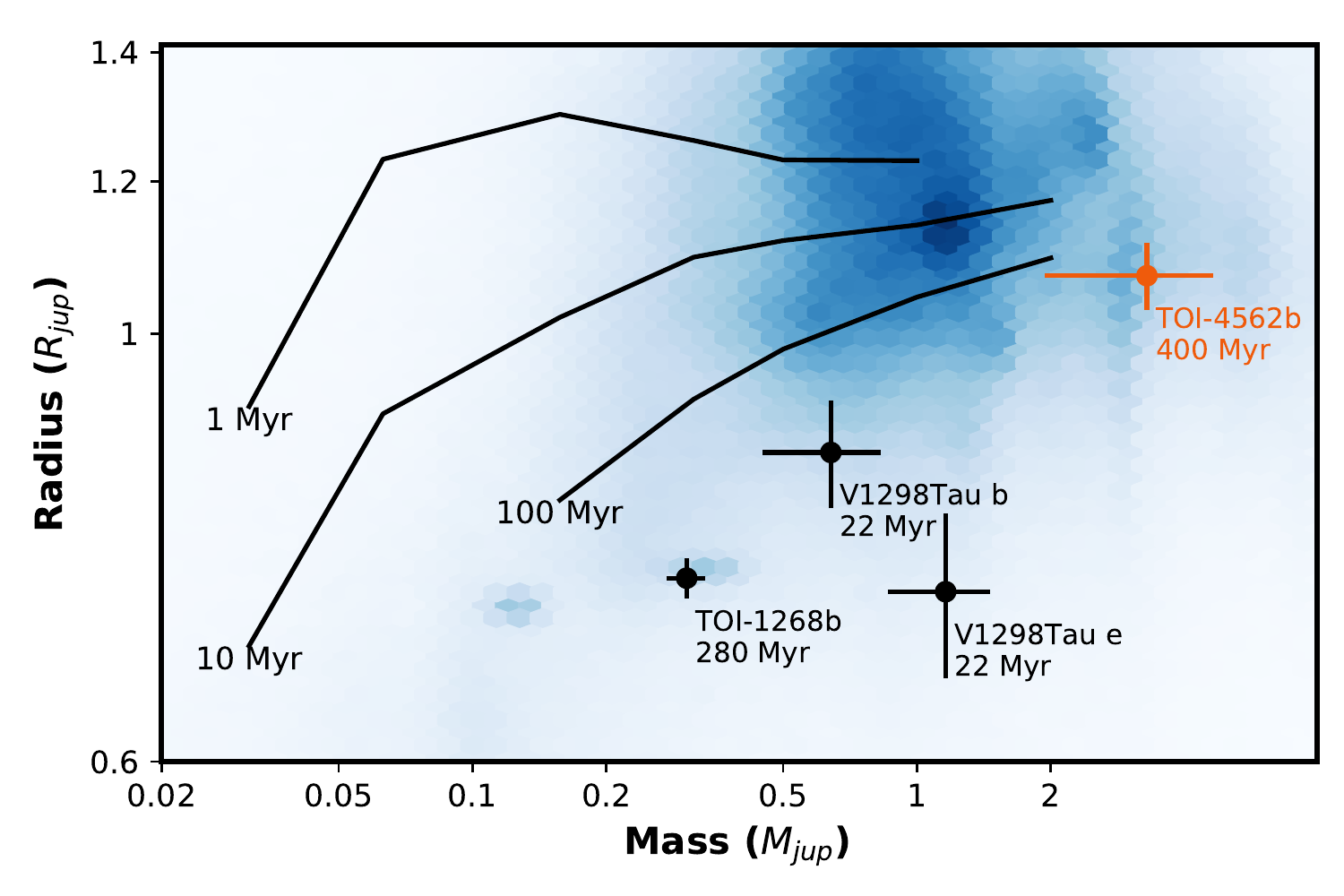}
    \caption{Young gas giants can help constrain the cooling and contraction models. To date, \toib is only the fourth Jovian planet younger than 500\,Myr to have both its mass and radius measured. The mass-radius of \toib is plotted in orange alongside the planets in the V1298 Tau \citep{2019ApJ...885L..12D,2021NatAs...6..232S} and TOI-1268 \citep{2022ApJ...926L...7D,2022arXiv220113341S} systems. Unlike others, \toib sits along the isochrone tracks that model the contraction of young planets \citep{2019A&A...623A..85L}. The mass-radius distribution of other known planets are shown with a density plot in blue in the background.}
    \label{fig:massradius}
\end{figure}

\subsection{Radius evolution}
\label{sec:contraction}

\begin{figure*}
    \centering
    \includegraphics[width=1\linewidth]{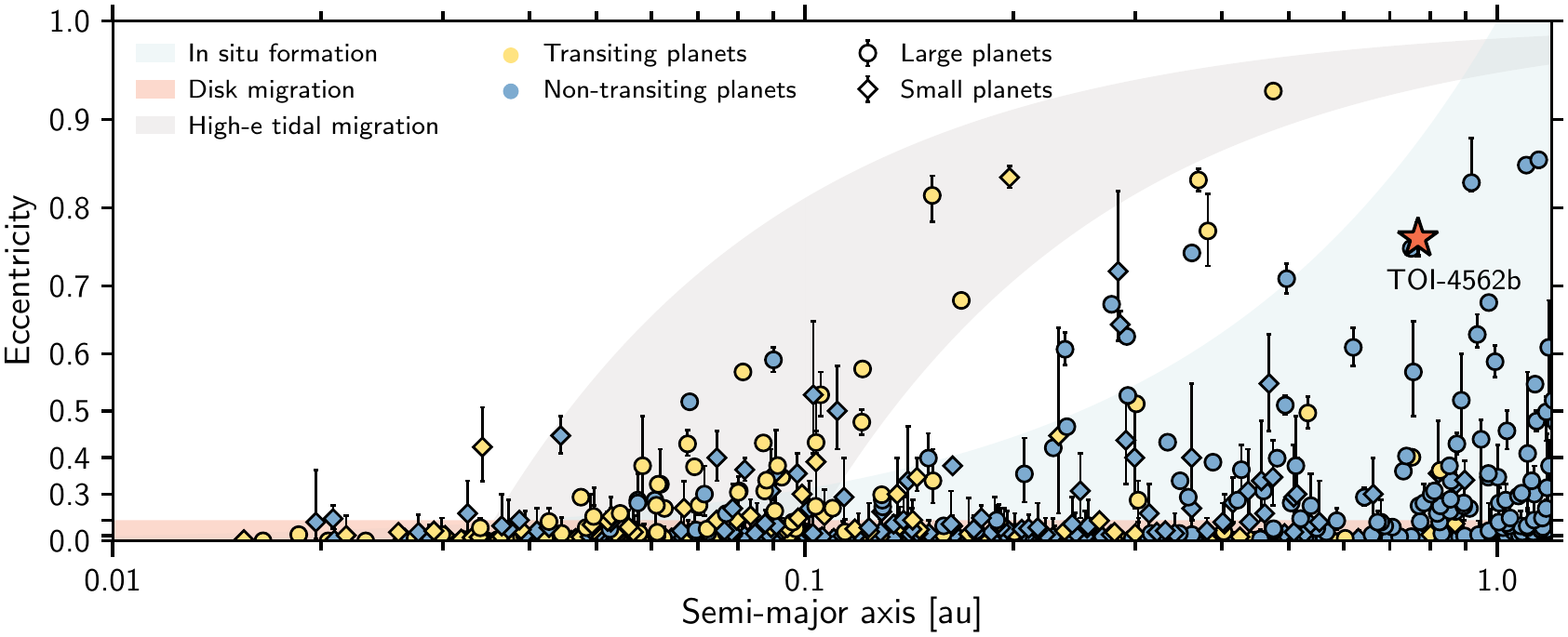}
    \caption{Eccentricity versus semi-major axis for all confirmed planets (Obtained from the NASA exoplanet archive 13 Feb. 2022) with \Mp\,$<$ 13\,\MJup. The vertical coordinate is scaled to $e^2$ to emphasize non-circular planets. Shaded areas highlight different formation scenarios. Planets in the grey region are on the path of high-eccentricity migration, with a final semi major axis between 0.034 and 0.1 au. The upper and lower bounds of this region are set by the Roche limit and the circularization timescale respectively. Disk migration, expected to only marginally excite orbital eccentricity is shown as the red shaded region. Finally, in situ formation, with eccentricity excited by, e.g., planet-planet scattering is shown in blue. Transiting versus non-transiting planets are labeled in yellow and blue respectively. Circles are representing larger (\Rp\,$>$\,6\,\REarth\,and/or \Mp\,$>$\, 100\,\MEarth) planets and diamonds smaller planets (\Rp\,$<$\,6\,\REarth\,and/or \Mp\,$<$\, 100\,\MEarth). Only planets with $e \geq$ 0.2 and with uncertainties on $e$ smaller than 50\% of the measured $e$ or planets with $e <$ 0.2 and with uncertainties less than 0.2 are shown.}\label{fig:ecc_sma}
\end{figure*}

\begin{figure}
    \centering
    \includegraphics[width=1\linewidth]{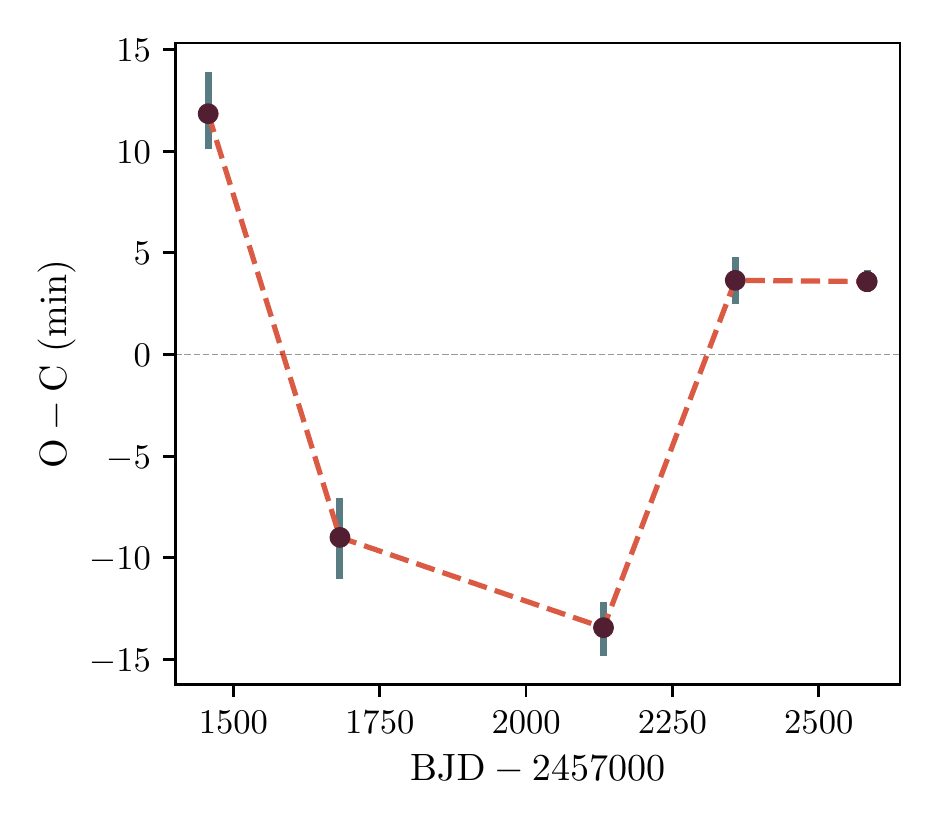}
    \caption{Observed - Calculated mid-transit time for the 5 transits of \toib, in minutes. The second and third transits (from TESS Sectors 13 and 30) show a $\sim$\,20 min mid-transit time difference with the other transits, suggesting the presence of a third body in the system.}
    \label{fig:ttv}
\end{figure}

At the end of their accretion phase, newly formed gas giants are expected to have radii larger than 1 \RJup. As the planet core radiates its primordial internal heat, Jovian mass planets will cool down via Kelvin-Helmholtz contraction to $\sim 1$~\RJup. Only Hot Jupiters, orbiting extremely close to their parent star
are expected to remain inflated due to their increased irradiation. 
According to cooling models \citep{2003A&A...402..701B,2007ApJ...659.1661F,2008A&A...482..315B,2019A&A...623A..85L}, shown on Figure~\ref{fig:massradius}, the most drastic changes in radius occur at the earliest ages. Measuring radii of young gas giants like \toib is therefore essential to set constraints on these such models, as emphasized in \cite{2007ApJ...659.1661F}.\par 

The current picture is unclear as the recently measured mass of V1298 Tau\,b\,\&\,e\, \citep{2021NatAs...6..232S} yield much denser planets than predicted at 20 Myr old and require dramatic heavy element enrichment to somewhat reconcile with cooling models (see Figure~\ref{fig:massradius}). Conversely, \toib's radius is as expected for its age. At the closest approach to its host star ($\sim0.18$\,AU), it receives stellar irradiation of $\sim$\,9.3\,$\times$\,10\,$^4$\,W m$^{-2}$, or $\sim$\,68 times that of Earth. Although above the $\sim$
\,1.6\,$\times$ 10\,$^4$\,W m$^{-2}$ threshold to trigger inflation, given by \cite{2018A&A...616A..76S} for planets more massive than 2.5~\MJup, \toib's orbital eccentricity means this level of irradiation affects the planet for a very short fraction of the orbit, not sufficient to trigger radius inflation. 

\begin{deluxetable*}{llll}
\tablenum{3}
\tablecaption{Next 10 transit opportunities for TOI-4562b.\label{tab:upcoming_transits}}
\tablewidth{0pt}
\tablehead{
\colhead{Transit mid-time ($\mathrm{BJD}$)} & \colhead{Transit date} & \colhead{Visible from (Partial (P) or Full (F))} & \colhead{TESS simultaneity}}
\decimals
\startdata
2460032.6977 & 29-Mar-2023 & Paranal (P) & Y     \\
2460257.8142 & 9-Nov-2023	& Paranal (P) & TBD   \\
2460482.9307 & 21-Jun-2024 & MKO (P) \& \textbf{ASTEP (F)} & TBD   \\
2460708.0473 & 1-Feb-2025	& \textbf{MKO (F)} & TBD   \\
2460933.1638 & 14-Sep-2025 & MKO (P) \& \textbf{ASTEP (F)} & TBD   \\
2461158.2803 & 27-Apr-2026 & SAAO (P) \& RUN (P) \& \textbf{ASTEP (F)} & TBD   \\
2461383.3968 & 8-Dec-2026	& SAAO (P) \& \textbf{RUN (F)} & TBD   \\
2461608.5134 & 27-Jul-2027 & ASTEP (P) & TBD   \\
2461833.6299 & 3-Mar-2028 & SAAO (P) \& \textbf{Paranal (F)} & TBD 	\\
\hline
\enddata
\tablenotetext{}{\textbf{Locations} SAAO: South African Astronomical Observatory, South Africa (latitude = -32.379444, longitude = -339.189306), Paranal: European Southern Observatory at Paranal, Chile (latitude = -24.625, longitude = -70.403333), MKO: Mt. Kent Observatory, Australia (latitude = -27.797861, longitude = 151.855417), RUN: Observatoire astronomique des Makes, Reunion Island (latitude = -21.199359, longitude = 55.409464), ASTEP: Antarctic Search for Transiting ExoPlanets, Dome C, Antarctica (latitude = -75.09978, longitude = 123.332196)}
\end{deluxetable*}

\subsection{Dynamical history of TOI-4562 b and benefits of additional follow-up} 
\label{subsec:evolution}

In its current observed state, \toib's semi major axis and eccentricity (see Figure~\ref{fig:ecc_sma}) are not in favour of a high eccentricity migration scenario as a circularization of its orbit would take orders of magnitudes longer than the age of the universe ($\tau_{circ} \sim$\,1\,$\times10^{7}$\,Gyr, \citealt{1966Icar....5..375G}). It is possible, however, that the planet is experiencing ongoing eccentricity cycles and we happen to be observing it at a lower eccentricity. Reduction of the star-planet distance at periastron at the eccentricity peak of such cycles might allow the circularization process to be triggered as described in \cite{2014ApJ...781L...5D}.
Disk-planet interactions can in principle excite the eccentricity of the orbit \citep{2015ApJ...812...94D} but this is restricted to low ($e \lesssim$\,0.2) values, as shown with the red area on Figure~\ref{fig:ecc_sma}. \cite{2021MNRAS.500.1621D} proposed that migration inside wide gaps carved in protoplanetary disks could result in gas giants with eccentricities up to 0.4. This is still insufficient to explain the very high eccentricity from \toib's orbit. \par
Another possible scenario to account for \toib's very high eccentricity is in-situ formation (or alternatively, smooth disk migration), followed by excitation from a companion. This can occur via secular interactions, or slow angular momentum exchanges with another body located further out, either periodically through e.g.,\,von Zeipel-Lidov-Kozai cycles \citep{1910AN....183..345V,1962P&SS....9..719L,1962AJ.....67..591K,2016ARA&A..54..441N,2008ApJ...678..498N} or chaotically in secular chaos \citep{2011ApJ...735..109W,Hamers2017}.
High eccentricity can also be triggered sporadically in planet-planet scattering \citep{1996Natur.384..619W,1996Sci...274..954R,2006ApJ...638L..45F,2008ApJ...686..580C}, or stellar fly-bys \citep{2016ApJ...816...59S,2021ApJ...913..104R}. Planet-planet scattering could have happened quickly and potentially early if triggered by the dissipation of the gas disk or if the planets were initially closely spaced. Constraints on an outer companion (if not ejected as a result of scattering) could provide crucial insights on dynamical evolution timescales give the young age of the system.\par 

The five transits of \toib show modest deviation from a linear ephemeris fit on the 5~--~20 min level (see Figure~\ref{fig:ttv}). This potential detection of a transit timing variation signal suggests the presence of a companion in the system, to which \toib probably owes its high eccentricity. The existing data are not sufficient to set meaningful constraints on the companion and most configurations for period (i.e., inner or outer companion), eccentricity and mutual inclination remain possible. \toib will be observed by \TESS again in its second extended mission in 2023. In Table~\ref{tab:upcoming_transits}, we show future opportunities to continue monitoring transits of \toib in the years to come. Combining these with long-term radial velocity follow-up might enable us to unravel the 3-D architecture and dynamical history of this system, as has been successfully performed for Kepler-419\,b\,\&\,c\,\citep{2012ApJ...761..163D,2014ApJ...791...89D}. We also note that no transit duration variations were found. \par
The orbital astrometric motion of an outer companion could be retrieved from \textit{Gaia} in the upcoming release of astrometric solutions for $\sim$~1.3 billion stars \citep{2021A&A...649A...2L}. When archival Hipparchos and \emph{Gaia} observations have been analysed jointly for previous brighter systems \citep[e.g.,][]{2021AJ....162...12V}, astrometric accelerations have often yielded constraints for outer stellar massed companions to key exoplanet systems. Additional \emph{Gaia} observations over the next $\sim 10$ years will allow us to achieve similar constraints for \toi. Combined with the diffraction limited adaptive optics observations estimated to reach $\sim$ 35 au (see section~\ref{subsec:direct-imaging}), these constraints can inform the presence of exterior stellar companions and provide means to distinguish between evolution scenarios.

Another candidate tracer for dynamical history is the angle between the star's rotation axis and the planet's orbital axis, or (sky projected) obliquity. From \Prot, \Rstar~and \vsini, we estimate the stellar inclination with respect to the line of sight to have a $3\sigma$ lower bound of $70^\circ$ as per \cite{2020AJ....159...81M}, consistent with being well aligned. Similarly to other planetary characteristics, the young ($<$\,1 Gyr) end of the obliquity distribution is under sampled. Recent measurements resulting from \TESS discoveries reveal a remarkable systematic alignment of young systems, including the Jupiter-sized planet HIP 67522 b \citep{2020AJ....160...33R,2021ApJ...922L...1H}, as well as a number of smaller planets (e.g., AU Mic b \& c; \cite{2020Natur.582..497P,2020A&A...643A..25P,2020A&A...641L...1M,2020ApJ...899L..13H, 2021AJ....162..137A}, DS Tuc Ab; \cite{2019ApJ...880L..17N,2020ApJ...892L..21Z,2020AJ....159..112M}, TOI 942 b \& c \cite{2021ApJ...917L..34W}, and TOI 251 \cite{2021AJ....161....2Z}).  
The estimated amplitude of the Rossiter McLaughlin effect \citep{1924ApJ....60...15R,1924ApJ....60...22M} for \toib is $\Delta$V\,$\sim$\,70--150\,\ms. Given the $\sim 4$ hours transit duration, combined with a brightness of \Vmag = 12.098 and a rotational broadening of $v$\,sin\,$i$ = 17.5\,\kms, this is well within the grasp of a 4m-class telescope and such an eccentric system would provide a precious addition to the age-obliquity distribution. It is important to note that the long orbital period remains a major obstacle to transit spectroscopy for ground-based facilities.

In the coming years, we aim to conduct extensive follow-ups of the \toi system to unravel the full architecture of the system and potentially provide insights into the processes shaping the current gas giant planet distribution. Such follow-up will include radial velocities, ground and space based photometry, astrometry and transit spectroscopy for obliquity measurements and/or atmospheric characterisation.

\begin{deluxetable*}{llll}
\tablenum{4}
\tablecaption{CHIRON radial velocities for \toi. The two left columns cover \toib's first orbit (late 2020 to mid 2021) and the two rights columns the second orbit (late 2021 to early 2022).\label{tab:RVs}}
\tablewidth{0pt}
\tablehead{
\colhead{$\mathrm{BJD}$} & \colhead{RV (\ms)} & \colhead{$\mathrm{BJD}$} & \colhead{RV (\ms)}}
\decimals
\startdata
2457919.279641 & 5362.7 $\pm$  78.4 & 2457951.584780 & 5312.4 $\pm$  84.9 \\
2457919.975844 & 5384.0 $\pm$ 107.5 & 2457951.882422 & 5298.0 $\pm$  99.8 \\
2457920.077478 & 5338.8 $\pm$  64.0 & 2457952.183280 & 5356.5 $\pm$  69.2 \\
2457920.770002 & 5472.3 $\pm$  80.6 & 2457952.483116 & 5341.9 $\pm$  81.1 \\
2457923.866595 & 5150.9 $\pm$  69.1 & 2457952.785734 & 5291.7 $\pm$ 116.4 \\
2457924.163829 & 5375.4 $\pm$  96.4 & 2457953.181198 & 5283.6 $\pm$  75.4 \\
2457926.463031 & 5361.0 $\pm$ 110.4 & 2457953.478271 & 5374.3 $\pm$  68.4 \\
2457926.769474 & 5386.8 $\pm$  75.5 & 2457953.779370 & 5412.8 $\pm$  62.9 \\
2457927.459646 & 5307.2 $\pm$  64.6 & 2457954.078933 & 5331.6 $\pm$  90.0 \\
2457927.755236 & 5311.9 $\pm$  93.8 & 2457954.477875 & 5593.4 $\pm$  68.6 \\
2457928.059467 & 5417.3 $\pm$  78.6 & 2457954.770163 & 5464.1 $\pm$  66.4 \\
2457928.466230 & 5204.1 $\pm$ 138.8 & 2457955.179212 & 5513.8 $\pm$  94.8 \\
2457928.761200 & 5199.1 $\pm$  80.0 & 2457955.280537 & 5461.7 $\pm$  89.2 \\
2457929.259193 & 5241.1 $\pm$  77.4 & 2457955.375845 & 5543.3 $\pm$  92.6 \\
2457929.558164 & 5370.1 $\pm$ 111.4 & 2457955.482781 & 5522.4 $\pm$  82.3 \\
2457929.953730 & 5152.4 $\pm$  83.3 & 2457955.578297 & 5313.6 $\pm$ 101.5 \\
2457930.253505 & 5155.4 $\pm$ 127.6 & 2457955.677910 & 5531.5 $\pm$  77.8 \\
2457930.457509 & 5435.7 $\pm$  74.4 & 2457955.978882 & 5386.7 $\pm$  76.3 \\
2457930.752931 & 5245.4 $\pm$ 135.4 & 2457956.078722 & 5357.4 $\pm$  49.4 \\
2457931.049893 & 5485.1 $\pm$  54.3 & 2457956.174936 & 5429.0 $\pm$  78.6 \\
2457931.459114 & 5335.7 $\pm$ 122.9 & 2457956.268873 & 5451.4 $\pm$ 128.8 \\
2457931.852818 & 5329.3 $\pm$  89.8 & 2457956.477261 & 5454.5 $\pm$  88.7 \\
2457932.354462 & 5475.9 $\pm$  87.3 & 2457956.567821 & 5509.4 $\pm$  74.6 \\
2457932.753588 & 4941.1 $\pm$ 191.1 & 2457956.671976 & 5411.0 $\pm$  75.6 \\
2457932.950353 & 5350.4 $\pm$  77.1 & 2457956.776594 & 5372.3 $\pm$  78.1 \\
2457933.157726 & 5182.6 $\pm$ 112.6 & 2457956.872101 & 5522.2 $\pm$ 121.3 \\
2457933.548574 & 5679.6 $\pm$ 112.6 & 2457956.968008 & 5465.6 $\pm$ 120.6 \\
2457934.251269 & 5352.0 $\pm$  77.5 & 2457957.070850 & 5581.7 $\pm$  80.9 \\
2457934.548758 & 5707.1 $\pm$  88.4 & 2457957.173735 & 5413.1 $\pm$  59.1 \\
2457934.949392 & 5777.3 $\pm$ 104.8 & 2457957.273394 & 5376.5 $\pm$  83.0 \\
2457935.148356 & 5684.9 $\pm$ 160.8 & 2457959.269741 & 5373.7 $\pm$  120.7\\
2457935.644737 & 5348.6 $\pm$  74.5 & 2457959.369361 & 5139.8 $\pm$  94.1 \\
2457935.744986 & 5126.8 $\pm$ 101.0 & 2457959.565821 & 5321.7 $\pm$  74.9 \\
2457936.045174 & 5124.1 $\pm$ 104.9 & 2457959.662741 & 5260.0 $\pm$  81.5 \\
2457936.146598 & 5314.9 $\pm$  84.3 & 2457959.765341 & 5220.0 $\pm$  60.7 \\
2457936.345446 & 5322.3 $\pm$ 119.3 & 2457959.861135 & 5383.7 $\pm$  87.1 \\
2457936.438155 & 5581.4 $\pm$ 114.3 & 2457959.969637 & 5298.7 $\pm$ 108.3 \\
2457936.545021 & 5093.0 $\pm$ 100.9 & 2457960.060741 & 5323.5 $\pm$  60.7 \\
2457936.646301 & 5255.5 $\pm$  74.4 & 2457960.162519 & 5330.1 $\pm$  87.7 \\
2457936.744909 & 5271.4 $\pm$  71.9 & 2457960.265488 & 5354.1 $\pm$  53.1 \\
2457936.846528 & 5428.8 $\pm$  89.1 & - & - \\ 
2457936.947656 & 5270.0 $\pm$  69.6 & - & - \\
2457937.046070 & 5415.1 $\pm$ 119.0 & - & - \\
2457937.145298 & 5446.5 $\pm$ 132.6 & - & - \\
\hline
\enddata
\end{deluxetable*}

We respectfully acknowledge the traditional custodians of the lands on which we conducted this research and throughout Australia. We recognize their continued cultural and spiritual connection to the land, waterways, cosmos and community. We pay our deepest respects to all Elders, present and emerging people of the Giabal, Jarowair and Kambuwal nations, upon whose lands the MINERVA-Australis facility at Mount Kent is located. This research has been supported by an Australian Government Research Training Program Scholarship.
GZ thanks the support of the ARC DECRA program DE210101893.
GZ, SQ thank the support of the \TESS Guest Investigator Program G03007.
CH thanks the support of the ARC DECRA program DE200101840.
EG gratefully acknowledges support from the David and Claudia Harding Foundation in the form of a Winton Exoplanet Fellowship.
This work was supported by an LSSTC Catalyst Fellowship awarded by LSST Corporation to TD with funding from the John Templeton Foundation grant ID \# 62192.
This research has used data from the CTIO/SMARTS 1.5m telescope, which is
operated as part of the SMARTS Consortium by \href{http://secure-web.cisco.com/1TL5nionOJJUGi7T0X_YvX7RLRwbVQl20QG7s4LKeK1vpFY8M3UHYMuONVvV2D2hxli_pMi4YkHdTYel4ogZ3sJWN4axM8-5IsyCIPeIj7BfVIBOvp9a8iRKv2IM-wTBpjGA3xxZcH5lT4FNKBIoEstyJEEyUYzEKbDQyL4T1LQSiukl5eTarVlkS9YJbHf_HrjiuXV1gM1uXr7gdIdCbZg4CfJa_N8Qw38oz0KhpJ74RZ0rIcyg3XKCc6-HCDjlBrMtX3cpMKa1Kcya1SxY0UxXY0WkwM0zGeXYUYfbkp1Ce6jIBY8Evcz-YcyODRE4QWMlPqSDV66bKv5F1R3-RrkcH91Y7INyFOP6qJfGJKLRFJT-KNphpqmNc4Pf7zLVOIBjCEKsANmt1XTtzQN5AIPwKf-F1qd4b6KCZrqjHZIA/http\%3A\%2F\%2Fwww.recons.org}{RECONS}
This work has made use of data from the European Space Agency (ESA) mission
{\it Gaia} (\url{https://www.cosmos.esa.int/gaia}), processed by the {\it Gaia}
Data Processing and Analysis Consortium (DPAC,
\url{https://www.cosmos.esa.int/web/gaia/dpac/consortium}). Funding for the DPAC
has been provided by national institutions, in particular the institutions
participating in the {\it Gaia} Multilateral Agreement.
This work makes use of observations from the LCOGT network. Part of the LCOGT telescope time was granted by NOIRLab through the Mid-Scale Innovations Program (MSIP). MSIP is funded by NSF.
This research has made use of the NASA Exoplanet
Archive, which is operated by the California Institute of Technology, under contract with the National Aeronautics and Space Administration under the Exoplanet Exploration Program. 
Funding for the \TESS mission is provided by NASA's Science Mission directorate. We acknowledge the use of public \TESS Alert data from pipelines at the \TESS Science Office and at the \TESS Science Processing Operations Center. This research has made use of the Exoplanet Follow-up Observation Program website, which is operated by the California Institute of Technology, under contract with the National Aeronautics and Space Administration under the Exoplanet Exploration Program. This paper includes data collected by the \TESS mission, which are publicly available from the Mikulski Archive for Space Telescopes (MAST).
Resources supporting this work were provided by the NASA High-End Computing (HEC) Program through the NASA Advanced Supercomputing (NAS) Division at Ames Research Center for the production of the SPOC data products.
The Center for Exoplanets and Habitable Worlds is supported by the Pennsylvania State University and the Eberly College of Science.
RB and MH acknowledge support from ANID – Millennium Science Initiative – ICN12\_009.
NE thanks everyone who takes part in the Planet Hunters TESS citizen science project, which contributes to finding new and exciting planetary systems.

\facility{TESS, Exoplanet Archive, CTIO 1.5\,m, LCOGT, Gemini:Zorro, CTIO SOAR, ESO 2.2\,m}

\software{\textsc{AstroImageJ}~\citep{2017AJ....153...77C}, \textsc{astropy}~\citep{2013A&A...558A..33A,2018AJ....156..123A},
\textsc{baffles}~\citep{2020ApJ...898...27S},
\textsc{batman}~\citep{2015PASP..127.1161K}, \textsc{celerite}~\citep{celerite},  \textsc{emcee}~\citep{2013PASP..125..306F}, \textsc{pyastronomy}~\citep{pya}, \textsc{comove} (\url{https://github.com/adamkraus/Comove}), \textsc{pyphot} (\url{https://mfouesneau.github.io/pyphot/}), \textsc{radvel}~\citep{2018PASP..130d4504F}, \textsc{scikit-learn}~\citep{scikit-learn}, \textsc{minimint} (\url{https://zenodo.org/record/4900576}), \textsc{numpy}~\citep{harris2020array}, \textsc{matplotlib}~\citep{Hunter:2007}, \textsc{astropy}~\citep{astropy:2013,astropy:2018}, \textsc{unred} (\url{https://github.com/pbrus/unredden-stars}), \textsc{pandas}~\citep{reback2020pandas}, \textsc{corner}~\citep{corner}}

\newpage
    
\bibliography{main}{}
\bibliographystyle{aasjournal}

\end{document}